\newcommand{\remove}[1]{}

\documentclass[11pt]{article}

\usepackage{amsmath}
\usepackage[breaklinks]{hyperref}  % comment out if you do not have it
\usepackage{graphicx,enumerate}
\usepackage{sariel,fullpage,verbatim,path}

\usepackage[normalem]{ulem}

\newcommand{\MTR}{\mathcal{M}}
\newcommand{\ropt}{r_{\mathrm{opt}}}

\newcommand{\Ball}{\mathbf{b}}

\newcommand{\Pin}{P_{\mathrm{in}}}
\newcommand{\Pout}{P_{\mathrm{out}}}

\newcommand{\Gin}{G_{\mathrm{in}}}
\newcommand{\Gout}{G_{\mathrm{out}}}

\newcommand{\Rel}[1]{\mathrm{Rel}(#1)}
\newcommand{\dRel}[1]{\overline{\mathrm{Rel}}(#1)}
\newcommand{\depth}{\overline{d}}

\newcommand{\Dist}{d}

\newcommand{\DstM}{{\Dist_\MTR}}

\newcommand{\rep}{\mathrm{rep}}

\newcommand{\Spread}{\Phi}

\newcommand{\nettree}{net-tree}
\newcommand{\Nettree}{Net-tree}

\DeclareMathOperator{\lca}{\textsf{lca}}

\newcommand{\polylog}{\mathrm{polylog}}
\newcommand{\poly}{\mathrm{poly}}

\newcommand{\Mparagraph}[1]{\paragraph{#1}}

\newcommand{\NetPermutAlg}{{\tt{NetPermut\-Alg}}}

\newcommand{\Diam}{\mathrm{diam}}
\newcommand{\rcurr}{r_{\mathrm{curr}}}

\newcommand{\NetC}{\mathcal{N}_C}
\newcommand{\GreedyCluster}{{\tt{GreedyCluster}}}

\newcommand{\mlevel}[1]{\ell({#1})}
\newcommand{\parent}{\overline{\mathrm{p}}}
\newcommand{\tbase}{\tau}

\newcommand{\caselab}[1]{\label{case:#1}}
\newcommand{\caseref}[1]{(\ref{case:#1})}

\newcommand{\NN}{\mathbb{N}}
\newcommand{\br}{\overline{r}} 

\newcommand{\WSPD}{\ensuremath{\mathrm{WSPD}}}
\newcommand{\WSPDProc}{\ensuremath{{\text{\textsf{genWSPD}}}}}

\newcommand{\sP}{\mathsf{P}}
\newcommand{\sS}{\mathsf{S}}
\newcommand{\sQ}{\mathsf{Q}}
\newcommand{\sAprx}{\overline{\kappa}}%{\widehat{\epsilon}}

\newcommand{\Root}{\mathrm{root}}

\newcommand{\CA}{\mathcal{A}}
\newcommand{\CB}{\mathcal{B}}
\newcommand{\CN}{\mathcal{N}}

\newcommand{\Partition}{\textsf{Partition}}

\newcommand{\hphi}{\widehat{\phi}}

\newcommand{\hx}{\widehat{x}}
\newcommand{\wu}{\widehat{u}}

\newcommand{\CD}{\mathcal{D}}

\begin{document}

\title{Fast Construction of Nets in Low Dimensional Metrics and Their
   Applications}
% \thanks{See \urlSarielPaper{04/lipschitz} for the most
%      recent version of this paper.}}

\author{Sariel Har-Peled\SarielThanks{Work on this paper was partially
      supported by a NSF CAREER award CCR-0132901.}
   \and 
   Manor Mendel%
   \thanks{University of Illinois and California Institute of Technology;
      \texttt{mendelma@gmail.com}.}
}

 \date{}

\maketitle

\begin{abstract}
    We present a near linear time algorithm for constructing
    hierarchical nets in finite metric spaces with constant doubling
    dimension.  This data-structure is then applied to obtain improved
    algorithms for the following problems: Approximate nearest
    neighbor search, well-separated pair decomposition, 
    spanner construction, compact
    representation scheme, doubling measure, and computation of the
    (approximate) Lipschitz constant of a function.  In all cases, the
    running (preprocessing) time is near-linear and the space being
    used is linear.
\end{abstract}

%%%%%%%%%%%%%%%%%%%%%%%%%%%%%%%%%%%%%%%%%%%%%%%%%%%%%%%%%%%%%%%%%%
%%%%%%%%%%%%%%%%%%%%%%%%%%%%%%%%%%%%%%%%%%%%%%%%%%%%%%%%%%%%%%%%%%

\section{Introduction}
\seclab{intro}

Given a data set, one frequently wants to manipulate it and compute
some properties of it quickly. For example, one would like to cluster
the data into similar clusters, or measure similarity of items in the
data, etc. One possible way to do this, is to define a distance
function (i.e., metric) on the data items, and perform the required
task using this metric. Unfortunately, in general, the metric might be
intrinsically complicated (``high dimensional''), and various
computational tasks on the data might require high time and space
complexity. This is known in the literature as ``the curse of
dimensionality''.

One approach that got considerable attention recently is to define a
notion of dimension on a finite metric space, and develop efficient
algorithms for this case.  One such concept is the notion of doubling
dimension \cite{a-pldr-83, h-lams-01,gkl-bgfld-03}.  The
\emph{doubling constant} of metric space $\MTR$ is the maximum, over
all balls $\Ball$ in the metric space $\MTR$, of the minimum number of
balls needed to cover $\Ball$, using balls with half the radius of
$\Ball$. The logarithm of the doubling constant is the \emph{doubling
   dimension} of the space.  The doubling dimension can be thought as
a generalization of the Euclidean dimension, as $\Re^d$ has
$\Theta(d)$ doubling dimension.  Furthermore, the doubling dimension
extends the notion of growth restricted metrics of Karger and Ruhl
\cite{kr-fnngr-02}.

Understanding the structure of such spaces (or similar notions), and
how to manipulate them efficiently received considerable
attention in the last few years \cite{c-nnqms-99,kr-fnngr-02,
gkl-bgfld-03,hkmr-nnng-04,kl-nnsap-04, kl-bbcnn-04,
t-beald-04}.  

The low doubling metric approach can be justified in two levels.
\begin{enumerate}
    \item Arguably, non-Euclidean, low (doubling) dimensional metric
    data appears in practice, and deserves an efficient algorithmic
    treatment.  Even high dimensional Euclidean data may have some low
    doubling dimension structure which make it amenable to this
    approach.

    This view seems to be shared by many recent algorithmic papers on
    doubling metrics, but it still awaits a convincing empirical
    and/or theoretical support.

    \item Even if one is only interested in questions on Euclidean
    point sets, it makes sense to strip the techniques being used to
    their bare essentials, obtaining better understanding of the
    problems and conceptually simpler solutions.
\end{enumerate}
More arguments along these lines can be found in \cite{c-nnqms-99},
where the author advocates this approach.

In general, it is impossible to directly apply algorithmic results
developed for fixed dimensional Euclidean space to doubling metrics,
since there exists doubling metrics that can not embedded in Hilbert
space with low distortion of the distances
\cite{s-nhlpg-96,l-pwwmc-02}.  Hence, some of the aforementioned works
apply notions and techniques from fixed dimensional Computational
Geometry and extend them to finite metric spaces.

In particular, Talwar \cite{t-beald-04} showed that one can extend the
notion of \emph{well-separated pairs decomposition} (\WSPD{}) of
\cite{ck-dmpsa-95} to spaces with low doubling dimension.
Specifically, he shows that for every set $P$ of $n$ points having
doubling dimension $\dim$, and every $\eps>0$, there exists \WSPD{},
with separation $1/\eps$ and $O( n \eps^{-O(\dim)} \log \Spread )$
pairs, where $\dim$ is the doubling dimension of the finite metric
space, and $\Spread$ is the \emph{spread} of the point set, which is
the ratio between the diameter of $P$ and the distance between the
closest pair of points in $P$.  This is weaker than the result of
Callahan and Kosaraju \cite{ck-dmpsa-95} for Euclidean space, which
does not depend on the spread of the point set.

Krauthgamer and Lee \cite{kl-nnsap-04} showed a data structure for
answering $(1+\eps)$-approximate nearest neighbor queries on point set
$P$ with spread $\Spread$.  Their data structure supports insertions
in $O(\log \Spread \log \log \Spread)$ time.  The preprocessing time
is $O( n \log \Spread \log \log \Spread)$ (this is by inserting the
points one by one), and the query time is $O( \log \Spread +
\eps^{-O(\dim)})$.  In $\Re^d$ for fixed $d$, one can answer such
queries in $O( \log \log (\Spread/\eps))$ time, using near linear
space, see \cite{h-rvdnl-01} and references therein (in fact, it is
possible to achieve constant query time using slightly larger storage
\cite{hm-ckmkm-04}).  
Note however, that the latter results strongly use the
Euclidean structure.  Recently, Krauthgamer and Lee \cite{kl-bbcnn-04}
overcame the restriction on the spread, presenting a data-structure
with nearly quadratic space, and logarithmic query time.

Underlining all those results, is the notion of hierarchical nets.
Intuitively, hierarchical nets are sequences of larger and larger
subsets of the underlining set $P$, such that in a given resolution,
there is a subset in this sequence that represents the structure of
$P$ well in this resolution (a formal definition is given in
\secref{preliminaries}). Currently, the known algorithms for
constructing those nets require running time which is quadratic in
$n$.

An alternative way for constructing those nets is by the clustering
algorithm of Gonzalez \cite{g-cmmid-85}. The algorithm of Gonzalez
computes $2$-approximate $k$-center clustering by repeatedly picking
the point furthest away from the current set of centers. Setting
$k=n$, this results in a permutation of the points in the metric
space. It is easy to verify that by taking different prefixes of this
permutation, one gets hierarchical nets for the metric. However, the
running time of Gonzalez algorithm in this case is still quadratic.
Although, in fixed dimensional Euclidean space the algorithm of Gonzalez
was improved to $O(n \log k)$ time by Feder and Greene
\cite{fg-oafac-88}, and linear time by Har-Peled \cite{h-cm-04}, those
algorithms require specifying $k$ in advance, and they do not generate
the permutation of the points, and as such they cannot be used in
this case.

\paragraph{Our results.}
In this paper, we present improved algorithms for the aforementioned
applications, having near linear preprocessing time and linear space.
We also remove the dependency on the spread.  As such, we (almost)
match the known results in computational geometry for low dimensional
Euclidean spaces.

We assume that the input is given via a black box that can compute the
distance between any two points in the metric space in constant time.
Since the matrix of all $\binom{n}{2}$ distances has quadratic size,
this means that in some sense our algorithms have sublinear running
time.  This is not entirely surprising since subquadratic time
algorithms exist for those problems in fixed dimensional Euclidean
space. Thus, our paper can be interpreted as further strengthening the
perceived connection between finite spaces of low doubling dimensions
and Euclidean space of low dimension.  Furthermore, we believe that
our algorithms for the well-separated pair decomposition and
approximate nearest neighbor are slightly cleaner and simpler than the
previous corresponding algorithms for the Euclidean case.

\Mparagraph{{\Nettree}.} %
In \secref{nets} we present a $2^{O(\dim)} n \log n$ expected time
randomized algorithm for constructing the hierarchical nets
data-structure, which we call \nettree{}.

\Mparagraph{Approximate Nearest Neighbor (ANN).} %
In \secref{nn:n:l:s} we show a new data-structure for
$(1+\eps)$-approximate nearest neighbor query.  The expected
preprocessing time is $2^{O(\dim)} n \log n$, the space used is
$2^{O(\dim)}n$, and the query time is $2^{O(\dim)} \log n+
\eps^{-O(\dim)}$.  The quality of approximation of the nearest
neighbor required is specified together with the query.

This query time is almost optimal in the oracle model since there are
examples of point sets in which the query time is $2^{\Omega(\dim)}
\log n $ \cite{kl-nnsap-04}, and examples in which the query time is
$\eps^{-\Omega(\dim)}$.%
\footnote{Consider the set $\mathbb{Z}_m^n$ with the $\ell_\infty$
   norm, where $m= \ceil{\eps^{-1}/2}$, $n=\ceil{dim}$.  Consider a
   query at point $q$ satisfying, $\exists x_0\in \mathbb{Z}_m^n$ such
   that $d(q,x_0)=m-1$, and $\forall x\in \mathbb{Z}_m^n$, $x\neq
   x_0\; \Rightarrow\; d(q,x)=m$. Since $x_0$ can be chosen in
   adversarial way any randomized $(1+\eps)$-ANN query algorithm would
   have to make $\Omega(m^n)$ distances queries before hitting $x_0$.}

Our result also matches the known results of Arya \etal{}
\cite{amnsw-oaann-98} in Euclidean settings. Furthermore, our result
improves over the recent work of Krauthgamer and Lee, which either
assumes bounded spread \cite{kl-nnsap-04}, or requires quadratic space
\cite{kl-bbcnn-04}.  The algorithms in
\cite{amnsw-oaann-98,kl-nnsap-04,kl-bbcnn-04} are deterministic, in
contrast to ours.

\Mparagraph{Well-Separated Pairs Decomposition (\WSPD{}).}  %
In \secref{WSPD}, we show that one can construct a $\eps^{-1}$
well-separated pairs decomposition of $P$, in near linear time. The
number of pairs is $n \eps^{-O(\dim)}$.  The size of the \WSPD{} is
tight as there are examples of metrics in which the size of the
\WSPD{} is $n \eps^{-\Omega(\dim)} $.  Our result improves over
Talwar's \cite{t-beald-04} work, and matches the results of Callahan
and Kosaraju \cite{ck-dmpsa-95} (these algorithms are deterministic,
though).

\Mparagraph{Spanners.} %
A $t$-Spanner of a metric is a sparse weighted graph whose vertices
are the metric's points, and in which the graph metric is
$t$-approximation to the original metric.  Spanners were first defined
and studied in \cite{ps-gs-89}.  Construction of $(1+\eps)$-spanners
for points in low dimensional Euclidean space is considered in
\cite{k-aceg-88,c-wspda-95}.  Using Callahan's technique
\cite{c-wspda-95}, the \WSPD{} construction also implies a near
linear-time construction of $(1+\eps)$-spanners having linear number
of edges for such metrics.  Independently of our work, Chan et.
al.~\cite{cgmz-hrbgm-05} show a construction of $(1+\eps)$-spanner for
doubling metrics with linear number of edges.  Their construction is
stronger in the sense that the degrees in their spanner graph are
bounded by constant. However, they do not specify a bound on the
running time of their construction.

\Mparagraph{Compact Representation Scheme (CRS).} %
In \secref{CRS}, we construct in near linear time, a data-structure of
linear size that can answer approximate distance queries between pairs
of points, in essentially constant time.  CRS were coined
``approximate distance oracles'' in \cite{tz-ado-01}.  Our result
extends recent results of Gudmunsson \etal{} \cite{glns-adogg-02,
   glns-ador-02} who showed the existence of CRS with similar
parameters for metrics that are ``nearly'' fixed-dimensional Euclidean
(which are sub-class of fixed doubling dimension metrics).  We also
mention in passing that our CRS technique can be applied to improve
and unify two recent results \cite{t-beald-04,s-date-05} on distance
labeling.

\Mparagraph{Doubling Measure.} %
A doubling measure $\mu$ is a measure on the metric space with the
property that for every $x\in P$ and $r>0$, the ratio
$\mu(\Ball(x,2r))/\mu(\Ball(x,r))$ is bounded, where $\Ball(x,r)=\{
y:\; d(x,y)\leq r\}$.  Vol$'$berg and Konyagin~\cite{vk-omdc-87}
proved that for finite metrics (and in fact for complete metrics
\cite{ls-ecdmscdm-98}) the existence of doubling measure is
quantitatively equivalent to the metric being doubling.  This measure
has found some recent algorithmic applications \cite{s-date-05}, and
we anticipate more applications.  Following the proof of
Wu~\cite{w-hddmms-98}, we present in \secref{doubling:measure} a near
linear time algorithm for constructing a doubling measure.

\Mparagraph{Lipschitz Constant of a Mapping.}%
In \secref{Lipschitz}, we study the problem of computing the
Lipschitz constant of a mapping $f:P \rightarrow B$.  In particular,
we show how using \WSPD{} it is possible to approximate the Lipschitz
constant of $f$ in near linear time (in $|P|$) when $P$ has constant
doubling dimension (and $B$ is an arbitrary metric). We also obtain
efficient exact algorithms, with near linear running time, for the
case where $P$ is a set of points in one or two dimensional
Euclidean space.

\Mparagraph{Computing the Doubling Dimension.}%
Although not stated explicitly in the sequel, we assume in
\secref{preliminaries} through \secref{Lipschitz} that the doubling
dimension of the given metric is either known a priori or given as part of the
input.  This assumption is removed in \secref{doubling-approximation}
where we remark that a constant approximation of the doubling
dimension of a given a metric $\MTR$ can be computed in $2^{O(\dim)} n
\log n$ time. It is therefore possible to execute the algorithms
of \secref{preliminaries} through \secref{Lipschitz} 
with the same asymptotic running time, using the approximation
of the doubling dimension
(In all the cases where the doubling dimension is needed, any upper
bound on it will do, with accordingly degraded running time).

\bigskip

Most of the algorithms in this paper are randomized. However, our use
of randomness is confined to \lemref{small:ball} 
(except for \secref{lipschitz-euclidean}). This means that the
algorithms always return the desired result, with bounds on the
\emph{expected} running time.  This also gives the same asymptotic
bound with constant probability, using Markov inequality.
Furthermore, in the ANN and CRS schema, randomness is only used in the
preprocessing, and the query algorithms are deterministic.
\lemref{small:ball} can be easily derandomized in $O(n^2)$ time, thus
giving $n^2 \text{polylog}(n)$ deterministic algorithms for all
problems discussed here.  We do not know whether a non-trivial
derandomization is possible.

%-----------------------------------------------------------------
%-----------------------------------------------------------------
%-----------------------------------------------------------------
%-----------------------------------------------------------------

\section{Preliminaries}
\seclab{preliminaries}

Denote by $\MTR$ a metric space, and $P$ a finite subset $P\subset
\MTR$.  The \emph{spread} of $P$, denoted by $\Spread(P)$, is the
ratio between the diameter of $P$, and the distance between the
closest pair of points in $P$.  For a point $p \in \MTR$ and a number
$r \geq 0$, we denote by $\Ball(p,r) = \brc{ q\in \MTR|\;
   \DstM(p,q)\leq r }$ the ball of radius $r$ around $p$. The
\emph{doubling constant} $\lambda$ of $P$ defined as the minimum over
$m\in \NN$ such that every ball $\Ball$ in $P$ can be covered by at
most $m$ balls of at most half the radius. The doubling dimension of
the metric space is defined as $\log_2 \lambda$. A slight variation of
the doubling constant is that any subset can be covered by $\lambda'$
subsets of at most half the diameter. It is not hard to see that
$\log_2 \lambda$ and $\log_2 \lambda'$ approximate each other up to a
factor of $2$.  Since we will ignore constant factors in the
dimension, these two definitions are interchangeable.  It is clear
that $\log_2\lambda'(P) \leq \log_2 \lambda'(\MTR)$, thus the doubling
dimension of $P$ is ``approximately'' at most that of $\MTR$.

A basic fact about $\lambda$ doubling metric $\MTR$ that will be used
repeatedly is that if $P\subset \MTR$ has spread at most $\Spread$, then
$|P| \leq \lambda^{O({\log_2 \Spread})}$.  

% We begin with a definition of the main object in this paper.

\subsection{Hierarchy of Nets}

An $r$-net in a metric space $\MTR$ is a subset ${\cal N} \subset
\MTR$ of points such that $\sup_{x\in \MTR} d_\MTR(x,\mathcal{N})\leq
r$, and $\inf_{x,y\in \mathcal{N};\; x\neq y} d_\MTR(x,y) \geq
r/\alpha$, for some constant $\alpha\geq 1$. $r$-nets are useful
``sparse'' object that approximately capture the geometry of the
metric space at scales larger than $3r$.  In this paper we will
heavily use the following notion of hierarchical nets.

\begin{defn}[\Nettree]
    Let $P\subset \MTR$ be a finite subset.  A \nettree{} of $P$ is a
    tree $T$ whose set of leaves is $P$.  We denote by $P_v\subset P$ the
    set of leaves in the subtree rooted at a vertex $v\in T$.  With
    each vertex $v$ associate a point $\rep_v \in P_v$.  Internal
    vertices have at least two children. Each vertex $v$ has a level
    $\mlevel{v} \in \ZZ \cup\brc{-\infty}$.  The levels satisfy
    $\mlevel{v}<\mlevel{\parent(v)}$, where $\parent(v)$ is the parent
    of $v$ in $T$.  The levels of the leaves are $-\infty$.  Let
    $\tbase$ be some large enough constant, say $\tbase=11$.
    
    We require the following properties from $T$:
    \begin{itemizeA}
        \item[] \textbf{Covering property:} 
        For every vertex $v\in T$,
        \[
        \Ball\pth{\rep_v, \tfrac{2\tbase}{\tbase-1} \cdot
           \tbase^{\mlevel{v}}} \supset P_v.
        \]
    
        \item[]\textbf{Packing property:} For every non-root vertex
        $v\in T$,
        \[
        \Ball\pth{\rep_v, \tfrac{\tbase-5}{2(\tbase-1)}\cdot
           \tbase^{\mlevel{\parent(v)}-1}} \bigcap P \subset P_v.
        \]
            
        \item[] \textbf{Inheritance property:} For every non-leaf
        vertex $u\in T$ there exists a child $v\in T$ of $u$ such that
        $\rep_u=\rep_v$.
    \end{itemizeA}

    \deflab{net:tree}
\end{defn}

The \nettree{} can be thought of as a representation of nets from
all scales in the following sense.

\begin{proposition}
    Given a \nettree{}, let
    \[
    \NetC(l) = \brc{ \rep_u \sep{ {\mlevel{u}}< l \leq
          {\mlevel{\parent(u)}} }}.
    \] 
    Then the points in $\NetC(l)$ are pairwise $\tbase^{l-1}/4$
    separated; that is, for any $p,q \in \NetC(l)$ we have $\DstM(p,q)
    \geq \tbase^{l-1}/4$.  In addition, $P \subseteq \cup_{p\in
       \NetC(l)} \Ball(p,4\cdot \tbase^l)$.

    \proplab{net:tree}
\end{proposition}
\begin{proof}
    Let $p,q \in \NetC(l)$, and let $u$ and $v$ be the corresponding
    nodes in the \nettree{}, respectively. Consider the balls $\Ball_p =
    \Ball\pth{p,r_p}$ and $\Ball_q = \Ball\pth{q,r_q}$, where $r_p =
    \tfrac{\tbase-5}{2(\tbase-1)}\cdot \tbase^{\mlevel{\parent(u)}-1}$
    and $r_q=\tfrac{\tbase-5}{2(\tbase-1)}\cdot
    \tbase^{\mlevel{\parent(v)}-1}$. The sets $\Ball_p \cap P$ and
    $\Ball_q \cap P $ are fully contained in $P_u$ and $P_v$
    respectively, by the definition of the \nettree{}. Since $u$
    and $v$ are on different branches of the \nettree{}, $P_u$
    and $P_v$ are disjoint. But
    then $\DstM(p,q) \geq \max \sbrc{r_p,r_q} \geq
    \tfrac{\tbase-5}{2(\tbase-1)}\cdot \tbase^{l-1} \geq
    \tbase^{l-1}/4$, by the definition of $\NetC(l)$ and since
    $\tau=11$.
    
    Similarly, consider the set of nodes $V_C(l) = \sbrc{ u \sepS{
          {\mlevel{u}}< l \leq {\mlevel{\parent(u)}} }}$ realizing
    $\NetC(l)$.  For any $v \in V_C(l)$, we have $P_v \subseteq
    \Ball\pth{\rep_v, \tfrac{2\tbase}{\tbase-1} \cdot
       \tbase^{\mlevel{v}}} \subseteq \Ball\pth{\rep_v, 2\tbase^l
       /(\tbase -1)}\subseteq \Ball\pth{\rep_v, \tbase^l }$, since
    $\tbase \geq 3$.  Thus, $P \subseteq \cup_{v\in V_C(l)}
    \Ball(\rep_v, \tbase^l) = \cup_{p\in \NetC(l)} \Ball(p,\tbase^l)$,
    as required.
\end{proof}

\medskip

Although $\NetC(\cdot)$ are quantitatively weaker nets compared with the
greedy approach,\footnote{We have made no attempt to optimize the
   ratio between the packing and covering radii, and the one reported
   here can be (substantially) improved.  However, some degradation in
   this ratio seems to be unavoidable.}  they are stronger in the sense
that the packing and the covering properties respect the hierarchical
structure of the \nettree{}.

The packing and covering properties easily imply that each vertex has
at most $\lambda^{O(1)}$ children.  \Nettree{}s are roughly equivalent
to compressed quadtrees \cite{amnsw-oaann-98}. The \Nettree{} is also
similar to the $\mathrm{sb}$ data-structure of Clarkson
\cite{c-nnsms-02}, but our analysis and guaranteed performance is new.

\subsection{The Computational Model.}
The model of computation we use is the ``unit cost floating-point word
RAM model''.  More precisely, for a given input consisting of
$\poly(n)$ real numbers at the range $[-\Spread,-\Spread^{-1}]
\cup[\Spread^{-1},\Spread]$, and given an accuracy parameter $t\in
\NN$, the RAM machine has words of length $O(\log n + \log\log \Spread
+t)$.  These words can accommodate floating point numbers from the set
\[ 
\brc{\pm (1+x)2^y \sep{ x\in[0,1], x2^{-t}\in \NN, y\in[-
      n^{O(1)}\log^{O(1)} \Spread, n^{O(1)} \log^{O(1)} \Spread] \cap \ZZ
   }},
\] 
and integers from the set $\brc{ -(2^t n\log
\Spread)^{O(1)},\ldots,0,\ldots, (2^t n\log \Spread)^{O(1)}}$.  For
simplicity, we assume that the input given in this way is
\emph{exact}.  All the problems discussed in this paper have  an
accuracy parameter $\eps>0$.  We assume that $\eps^{O(1)}>2^{-t}$, to
avoid rounding problems.  The space used by an algorithm (or a scheme)
is the number of \emph{words} being used. The machine allows
arithmetic, floor, ceiling, conversion from integer to floating point,
logarithm and exponent operations in unit time.  We further assume
that the machine is equipped with a random number generator.

Floating-point computation is a very well studied topic, see
\cite[Ch.~4]{k-taocp:sa-97} and references therein.  However, we were
unable to trace a citation that explicitly defines an asymptotic
floating-point computational model.  We choose this model for two
related reasons:
\begin{enumerate}
    \item The algorithms in this paper are supposed to output only
    approximate solution.  Therefore it makes sense to try and use
    approximate numbers since they use less resources.
    
    \item An important theme in this paper is developing algorithms
    that are independent of the spread of the given metrics. 
    Most algorithms that have an explicit dependence on the spread in
    their time or space complexity, have some form of $\polylog( \Spread)$
    dependence. 
    An algorithm that has no dependence on the spread
    $\Spread$, but relies on words of length $O(\log \Spread)$, may be
    considered suspicious at best.
\end{enumerate}

Having said that, for the most part in the sequel we will ignore
numerical and accuracy issues in our algorithms.  The algorithms are
simple enough that it is evidently clear that no numerical stability
issues arise. A notable exception is Assouad's embedding discussed in
\secref{do:Assouad}. There we have to explicitly add another
ingredient (\lemref{red-poly}) to the algorithm in order to adapt it
to the floating point word RAM model.  Indeed, that section is the
catalyst for the current discussion.

\subsection{Finding a Separating Ring}

We next present a simple argument that helps to overcome
the dependence on the spread in the running time.

\begin{observation}
Denote by $\ropt(P, m)$ the radius of
    the smallest ball in $P$ (whose center is also in $P$) 
    containing $m$ points.
    Then in a metric space with doubling constant $\lambda$,
      any ball of radius $2r$, where $r \leq 2\ropt(P, m)$, 
    contains at most $\lambda^2 m$ points. 
\end{observation}

\begin{proof}
    By the doubling property, the ball of radius $2r$ can be covered by
    $\lambda^2$ balls of radius $\ropt(P, m)$. Each such ball contains
    at most $m$ points.
\end{proof}

\begin{lemma} 
    Given an $n$-point metric space $P$ 
    with doubling constant $\lambda$, one can compute a ball
    $\Ball = \Ball(p,r)$, such that $\Ball$ contains at least $m = n/(
    2\lambda^3)$ points of $P$, and $\Ball(p,2r)$ contains at most
    $n/2$ points of $P$. The expected running time of this algorithm
    is $O(\lambda^3 n)$.

    \lemlab{small:ball}
\end{lemma}

\begin{proof}
    Pick randomly a point $p$ from $P$, and compute the ball
    $\Ball(p,r)$ of smallest radius around $p$ 
    containing at least $n/(2\lambda^3)$
    points. Next, consider the ball of radius $\Ball(p,2r)$. If it
    contains $\leq n/2$ points we are done. Otherwise, we repeat this
    procedure until success.
    
    To see why this algorithm succeeds with constant probability in
    each iteration, consider the smallest ball
    $Q = P \cap \Ball(q,\ropt)$ that contains at least $m$ points of
    $P$. Observe that any ball of radius $\ropt/2$ contain less than
    $m$ points.  With probability $1/(2\lambda^3)$ our sample is from
    $Q$. If $p \in Q$, then $r \leq 2\ropt$, and by the doubling
    property the ball $\Ball( p, 4\ropt )$ can be covered by at most
    $\lambda^3$ balls of radius $\ropt/2$. Hence it holds that
    $\cardin{P \cap \Ball(p,2r)} < \lambda^3 m \leq n/2$.
    
    Thus, the algorithm succeeds with probability $1/(2\lambda^3)$ in
    each iteration, and with probability $\geq 1/3$ after $2\lambda^3$
    iterations, implying the result, as each iteration takes 
    $O(n)$ time.
\end{proof}

\lemref{small:ball} enable us to find a sparse ring of radius ``not
much larger'' than its width.  For example, using it we can find an
empty ring of width $h$ and radius at most $2n h$ in linear time.

%-------------------------------------------------------------------------
%-------------------------------------------------------------------------

\section{Computing Nets Efficiently}
\seclab{nets}

In this section we prove the following theorem.

\begin{theorem}
    Given a set $P$ of $n$ points in $\MTR$, one
    can construct a \nettree{} for $P$ in $2^{O(\dim)}n \log n $ expected time.

    \theolab{net:tree} 
\end{theorem}

The outline of the proof is as follows. In
\secref{Gonzalez:low:spread} we show how to construct Gonzalez
sequence in $2^{O(\dim)}n \log (n+\Spread)$ time.  We then eliminate
the dependence of the running time on the spread $\Spread$, in
\secref{Gonzalez:high:spread}, using a tool developed in \secref{HST}.
In \secref{net:tree} we conclude the proof of \theoref{net:tree} by
showing how to construct the {\nettree} from the Gonzalez sequence.
We end with mentioning in \secref{aug} few data structures for
efficient searching on the \nettree{}.

\subsection{Computing greedy clustering quickly}
\seclab{Gonzalez:low:spread}

Gonzalez \cite{g-cmmid-85} presented a greedy algorithm, denoted by
\GreedyCluster{}, that when applied to a set of points $P$, computes a
permutation of the points $\Pi = \permut{p_1, p_2, \ldots, p_m}$, such
that $p_1, \ldots, p_k$ are good centers for $P$, for any $k \geq 1$.
We refer to $\Pi$ as the \emph{greedy permutation} of $P$.  Formally,
there are numbers $r_1, \ldots, r_n$, such that $P \subseteq
\cup_{l=1}^{k} \Ball(p_l, r_k)$.  Furthermore, $\min_{1 \leq i < j
   \leq k} d_\MTR(p_i, p_j) = r_{k-1}$.

\GreedyCluster{} works by picking an arbitrary point in $P$ to be
$p_1$, and setting $r_1$ to be the distance of the furthest point in
$P$ to $p_1$. \GreedyCluster{} stores for every point $q \in P$ its
distance to the closest center picked so far; namely, in the beginning
of the $k$th iteration, for all $q \in P$ we have $\alpha_q^k =
\min_{i=1}^{k-1} d_\MTR(q, p_i)$. The algorithm sets the $k$th center
to be $p_k = \arg \max_{p \in P} \alpha_p^k$ (namely, $p_k$ is the
point in $P$ furthest away from the centers picked so far). Clearly,
$r_{k-1} = \alpha_{p_k}^k$.  Implemented naively, one can compute the
first $k$ points $p_1, \ldots, p_k$ in $O(n k)$ time.  This requires
just scanning the points $k$ times. In the $k$th iteration, updating
$\alpha_q^k = \min( \alpha_q^{k-1}, d_\MTR(q, p_{k-1}))$, and
computing the point with the maximum such value. Thus, this leads to a
$2$-approximation to $k$-center clustering in $O(n k)$ time.

Feder and Greene \cite{fg-oafac-88} improved the running time to $O(n
\log {k})$ time (this was further improved to linear time by Har-Peled
\cite{h-cm-04}). Feder and Greene's main observation was that when
updating $\alpha_q^{k+1}$ one needs to update this value only for
points of $P$ which are in distance $\leq r_{k-1}$ away from $p_{k}$,
since for points $q$ further away, the addition of $p_k$ can not
change $\alpha_q^k$.

This suggests a natural approach for computing the greedy permutation:
Associate with each center in $\{p_1, \ldots, p_k\}$ the points of $P$
that it serves (namely, points that are closer to the given center
than to any other center). Furthermore, each center $p_i$, maintains a
\emph{friends list} that contains all the centers that are in distance
at most $4 r_k$ from it.  An ``old'' center will trim a point from its
friends list only when it its distance is larger than $8r_k$.
Specifically, the friends list of $p_i$ at the $k$th iteration ($k\geq
i$) contains all the centers at distance at most $\min\{8r_k, 4r_i\}$
from $p_i$.  Because of the constant doubling dimension property, this
list is of size $\lambda^{O(1)}$.

We further maintains a max-heap in which every center $p_i$, $i< k$
maintains the point $p'_i$ furthest away from $p_i$ in its cluster
along with its current $\alpha_{p'_i} = \DstM(p_i,p'_i)$ value.

At the $k$th iteration, the algorithm extracts the maximum value from
the heap. It sets $p_k$ to be the corresponding point. Denote by
$c_{p_k}$ the closest point among $\{p_1,\ldots, p_{k-1}\}$ to $p_k$
(i.e., the cluster's center of $p_k$ at the end of the $(k-1)$th round).
Next, the algorithm scans all the points currently served by the same
cluster as $c_{p_k}$, or by clusters containing points from friends
list of $c_{p_k}$, and update the $\alpha$ value of those points.
Furthermore, it moves all the relevant points to the newly formed
cluster. In the process, it also update the points $p'_i$ (of maximum
distance from $p_i$ in its cluster) for all $p_i$ in the friends list
of $c_{p_k}$.  It also computes friends list of $p_k$ (how to exactly
do it will be described in detail shortly).

We next bound the running time. To this end, a phase starting at the
$i$th iteration of the algorithm terminates at the first $j > i$ such
that $r_{j-1} \leq r_{i-1}/2$.  A ball of radius $4r_{j-1}$ around
each point $q \in P$ contains at most $\lambda^3$ points of $p_1,
\ldots, p_j$, and as such every point of $P$ is being scanned at most
$\lambda^3$ times at each phase of the algorithm. Thus, if the spread
of the point set is $\Spread$, the number of phases is $O(\log
\Spread)$, and scanning takes $\lambda^{O(1)} n \log \Spread $ time
overall.  Maintaining the max-heap costs an additional $
\lambda^{O(1)} n \log n$ time, since in each iteration only
$\lambda^{O(1)}$ values in the head are changed.

The only remaining hurdle is the computation of the friends list of a
newly formed center $p_k$. This can be done by maintaining for every
point $p_{l}$, $l\in \{1,\ldots,n\}$, the serving center $p_{l'}$ two
phases ago (at the end of that phase).  The new friends list of
$p_k$ can be constructed by scanning the friends list
of $p_{k'}$, and picking those in $\Ball(p_k,4r_k)$.  
This costs $\lambda^{O(1)}$ time for $p_k$ and
$O(\lambda^{O(1)} n)$ time overall.  
To see that this search suffices, we should see
that the set $\{p_i| i<k,\; \DstM(p_i,p_k)\leq 4r_k\}$ is scanned.
Indeed, fix $p_{i_0}$, having $i_0<k$, and $\DstM(p_{i_0},p_k)\leq
4r_k$.  Let $p_{k'}$ be the center of $p_k$ two phases ago. From the
definition, $2r_k\leq r_{k'} \leq 4r_k$, so $\DstM(p_k,p_{k'})\leq
4r_k$.  The current (at the end of the $(k-1)$th iteration) 
friends list of $p_{k'}$ contains all the
current centers at distance at
most $\min\{8r_k, 4r_{k'}\}=8r_k$ from $p_{k'}$. Furthermore,
\[
\DstM(p_{i_0},p_{k'})
\leq \DstM(p_{i_0}, p_k)+ \DstM(p_k,p_{k'})\leq 8r_k.
\]
we are therefore guaranteed that $p_{i_0}$ will be scanned.

Of course, as the algorithm progresses it needs to remove non-relevant
elements from the friends list as the current clustering radius $r_i$
shrinks. However, this can be done in a lazy fashion whenever the
algorithm scans such a list.

\begin{theorem}
    \theolab{h:nets} Let $P$ be a $n$-point metric space with doubling
    constant $\lambda$ and spread $\Spread$. Then the greedy
    permutation for $P$ can be computed in $O(\lambda^{O(1)} n \log
    (\Spread n))$ time, and $O(\lambda^{O(1)} n)$ space.
\end{theorem}

% ---------------------------------------

\subsection{Low Quality Approximation by HST}
\seclab{HST}

Here we present an auxiliary tool that will be used in
\secref{Gonzalez:high:spread} to extend the {\nettree} construction of
\secref{Gonzalez:low:spread} to metric spaces with large spread.

We will use the following special type of metric spaces:
\begin{defn}
    \emph{Hierarchically well-separated tree} (HST) is a metric space
    defined on the leaves of a rooted tree $T$. With each vertex $u\in
    T$ there is associated a label $\Delta_u \ge 0$ such that
    $\Delta_u=0$ if and only if $u$ is a leaf of $T$.  The labels are
    such that if a vertex $u$ is a child of a vertex $v$ then
    $\Delta_u\leq \Delta_v$. The distance between two leaves $x$ and
    $y$ of $T$ is defined as $\Delta_{\lca(x,y)}$, where $\lca(x,y)$
    is the least common ancestor of $x$ and $y$ in $T$.

    \deflab{HST}
\end{defn}

The class of HSTs coincides with the class of finite ultrametrics.
For convenience, we will assume that the underlying tree is binary
(any HST can be converted to binary HST in linear time, while
retaining the underlying metric).  We will also associate with every
vertex $u\in T$, an arbitrary leaf $\rep_u$ of the subtree rooted at
$u$. We also require that $\rep_u \in \brc{\rep_v |\; v \text{ is a
         child of } u}$.

A metric $N$ is called $t$-approximation of the metric $\MTR$, if they are
on the same set of points, and $\DstM(u,v)\leq d_N(u,v) \leq t \cdot
\DstM(u,v)$, for any $u,v \in \MTR$.

It is not hard to see that any $n$-point metric is
$(n-1)$-approximated by some HST (see, e.g. \lemref{Kruskal}). Here we
show:

\begin{lemma}
    For $n$-point metric space $\MTR$ with constant doubling dimension, it is
    possible to construct in $O( n \log
    n)$ expected time an HST which is $3n^2$ approximation of $\MTR$.

    \lemlab{l:q:HST}
\end{lemma}

This low quality HST will help us later in eliminating the dependence
on the spread of the construction time the \nettree{} and in distance
queries.

We begin proving \lemref{l:q:HST},
by constructing a sparse graph that approximates the original
metric  (this is sometimes called \emph{spanner}).

\begin{lemma} 
    Given an $n$-point metric space $P$ with doubling constant $\lambda$,
    one can compute a weighted graph $G$ that $3n$-approximates $P$ in
    $O(\lambda^6 n\log n )$ expected time.  
    The graph $G$ contains $O(\lambda^3 n \log n)$ edges.

    \lemlab{l:q:spanner}
\end{lemma}
\begin{proof}
    The construction is recursive. If $n = O(1)$, we just add all the
    pairs from $P$ as edges. Otherwise, we compute, using
    \lemref{small:ball}, a ball $\Ball(c,r)$ containing at least $m =
    n/( 2\lambda^3)$ points of $P$ with the additional property that
    $\Ball(c,2r)$ contains at most $n/2$ points of $P$.
    
    As such, there exists two numbers $r', h$ such that $r\leq r' \leq
    2r$, $h \geq r/n$ and $P \cap \Ball(c,r') = P \cap \Ball(c,r' +
    h)$ (namely, the ring with outer radius $r' + h$ and inner radius
    $r'$ around $c$ is empty of points of $P$). Computing $r'$ and $h$
    is done by computing the distance of each point from $c$, and
    partitioning the distance range $[r,2r]$ into $2n$ equal length
    segments. In each segment, we register the point with minimum and
    maximum distance from $c$ in this range. This can be easily done
    in $O(n)$ time using the floor function. Next, scan those buckets
    from left to right. Clearly, the maximum length gap is realized by
    a maximum of one bucket together with a consecutive non empty
    minimum of another bucket. Thus, the maximum length interval can
    be computed in linear time, and yield $r$ and $h$.

    Let $\Pin = \Ball(c,r') \cap P$ and let $\Pout = P \setminus
    \Pin$. Observe that $d_\MTR(\Pin, \Pout)= \min_{p \in \Pin, q \in
       \Pout}$ $d_\MTR(p,q) \geq h \geq r/n$. Next, we build recursively
    a spanner for $\Pin$ and a spanner for $\Pout$. We then add
    the edges between $c$ and all the points of $P$ to the spanner.
    Let $G$ denote the resulting graph.
    
    Since $n/2 \geq \cardin{\Pin} \geq n/ 2\lambda^3$ points of $P$,
    the running time of the algorithm is $T\pth{\cardin{P}} =
    T\pth{\cardin{\Pin} } + T\pth{ \cardin{\Pout}} + O\pth{\lambda^3
       n} = O(\lambda^6 n \log n)$. Similarly, the number of edges in
    $G$ is $O( \lambda^3 n \log n)$.
    
    We remain with the delightful task of proving that $G$ provides a
    $3n$-approximation to the distances of $P$. Let $\Gin$ and $\Gout$
    be the the graphs computed for $\Pin$ and $\Pout$, respectively.
    Consider any two points $u,v\in P$.  If $u$ and $v$ are both in
    $\Pin$ or both in $\Pout$ then the claim follows by induction.
    Thus, consider the case that $u \in \Pin$ and $v \in \Pout$.
    Observe that $d_\MTR(u,v) \geq h \geq r/n$. On the other hand,
    \[
    r/n \leq d_\MTR(u,v) \leq d_G(u,v) \leq d_\MTR(c,u) + d_\MTR(c,v) \leq
    r + r + d_\MTR(u,v) \leq (2n+1) d_\MTR(u,v),
    \]
    since $d_\MTR(c,v) \leq d_\MTR(c,u) + d_\MTR(u,v) \leq r +d_\MTR(u,v)$.
    Clearly, this implies that $d_G(u,v) \leq 3n d_\MTR(u,v)$, as
    claimed.    
\end{proof}

We will later obtain in \theoref{spanner} a near linear time
construction of spanners that $(1+\eps)$-approximate the original
metric and have linear number of edges.

\begin{lemma}
    Given a weighted connected graph $G$ on $n$ vertices and $m$
    edges, it is possible to construct in $O(n \log n + m)$ time an
    HST $H$ that $(n-1)$-approximates the shortest path metric on $G$.

    \lemlab{Kruskal}
\end{lemma}
\begin{proof}
    Compute the minimum spanning tree of $G$ in $O(n \log n + m)$
    time, and let $T$ denote this tree.
    
    Sort the edges of $T$ in non-decreasing order, and add them to the
    graph one by one. The HST is built bottom up. At each point we
    have a collection of HSTs, each corresponds to a connected
    component of the current graph. When an added edge merges two
    connected components, we merge the two corresponding HSTs into one
    by adding a new common root for the two HST, and labeling this
    root with the edge's weight times $n-1$. This algorithm is only
    a slight variation on Kruskal algorithm, and has the same running
    time.
    
    We next estimate the approximation factor. Let $x$ and $y$ be two
    vertices of $G$. Denote by $e$ the first edge that was added in
    the process above that made $x$ and $y$ to be in the same
    connected component $C$. Note that at that point of time $e$ is
    the heaviest edge in $C$, so $w(e)\leq d_G(x,y)\leq (|C|-1)\, w(e)
    \leq (n-1) \, w(e)$.  Since $d_H(x,y)=(n-1)\, w(e)$, we are done.
\end{proof}

The proof \lemref{l:q:HST} now follows by applying \lemref{Kruskal}
on the spanner from \lemref{l:q:spanner}.

Note that by applying \lemref{Kruskal} on the spanner from
\theoref{spanner},
one can obtain a near linear
time construction of HST which $O(n)$ approximates that original
metric.

%-----------------------------------------------------
%-----------------------------------------------------

\subsection{Extending greedy clustering to metrics of large spread}
\seclab{Gonzalez:high:spread}

The main idea in removing the dependence of the running time on the
spread is to apply the algorithm of \secref{Gonzalez:low:spread} to a
\emph{dynamic} set of points that will correspond to a level of the
HST.  In more details, the set of points will correspond to the
representatives $\rep_v$, where $\Delta_v \leq \rcurr/ n^4 \leq
\Delta_{\parent(v)} $, where $\rcurr$ is the current greedy radius,
$\Delta_v$ is the HST label of $v$ (i.e. the diameter of subtree
rooted at $v$), and $\parent(v)$ is the parent of $v$ in the HST.  The
algorithm now needs to handle another type of event, since as the
algorithm proceeds, the greedy radius decreases to a level in which
$\Delta_v \geq \rcurr/n^4$. In this case $v$ should be replaced by its
two children $u, w$.  Specifically, if $v$ belongs to a cluster of a
point $p_i$, we remove $\rep_v$ from the list of points associated
with the cluster of $p_i$, and add $\rep_{u}$ and $\rep_{w}$ to this
list (the case where $p_i$ is equal to $\rep_v$ is handled in a
similar fashion). Next, we need to compute for the new point its
nearest center; namely, compute $\alpha_{\rep_u}$ and
$\alpha_{\rep_w}$ (in fact, since $\rep_v = \rep_{u}$ or $\rep_v =
\rep_{w}$, we need to compute only one of those values). To this end,
we scan the friend list of $p_i$, and compute $\alpha_{\rep_u}$ and
$\alpha_{\rep_w}$ from it.  This takes $\lambda^{O(1)}$ time. We also
need to insert $\brc{\rep_u,\rep_w} \setminus \brc{\rep_v}$ into the
max-heap.

Thus, the algorithm has two heaps. One, is a max-heap maintaining the
points according to their distances to the nearest center, that is for
every point $p \in P$ we maintain the values of $\alpha_p$ in a
max-heap. The second max-heap, maintains the nodes of the HST sorted
by their diameters $\Delta$ (multiplied by a factor of $n^{4}$
for normalization). At every point, the algorithm extract the
larger out of two heaps, and handle it accordingly. One important
technicality, is that the algorithm is no longer generating the same
permutation as \GreedyCluster{}, since we are not always picking the
furthest point to add as the next center. Rather, we add the furthest
active point. We refer to the new algorithm as \NetPermutAlg{}.

\begin{lemma}
    Let $\pi = \permut{p_1, \ldots, p_n}$ be the permutation of $P$
    generated by \NetPermutAlg{}. Furthermore, let $r_k =
    \alpha_{p_{k+1}}^{k+1} = \min_{i=1}^{k} d_\MTR(q, p_i)$. Then, $P
    \subseteq \cup_{i=1}^{k} \Ball(p_i, (1+n^{-2}) r_k)$ and for any
    $u,v \in \brc{p_1,\ldots, p_k}$ we have $\DstM(u,v) \geq
    (1-n^{-2})r_k$.

    \lemlab{g:weak}
\end{lemma}

\begin{proof}
    Clearly, the  ball of radius $r_k$ around $p_1,\ldots, p_k$ cover
    all the active points when $p_{k+1}$ was created. However, every
    active point might represent points which are in distance
    $r_k/n^2$ from it. Thus, by expanding the radius by $(1+1/n^2)$,
    those balls cover all the points.
    
    Observe, that this implies that for any $i <j$ we have
    $(1+n^{-2})r_i \geq r_j$. In particular, let $\alpha \leq k$ be
    the minimum number such that $u,v \in \brc{p_1,\ldots, p_\alpha}$.
    Clearly, $\DstM( u,v) \geq r_{\alpha -1} \geq r_k /(1+n^{-2})
    \geq (1-n^{-2}) r_k$.
\end{proof}

\begin{lemma}
    The expected running time of \NetPermutAlg{} is 
    $O(\lambda^{O(1)}n \log n)$.
\end{lemma}

\begin{proof}
    Constructing the HST takes $\lambda^{O(1)}n\log n $ expected time, using
    \lemref{l:q:HST}. As in the bounded spread case, we conceptually
    divide the execution of the algorithm into phases. In the $i$th
    phase, the algorithm handles new clusters with radii in the range
    $\Diam(P)/2^{i-1}$ and $\Diam(P)/2^i$.  Consider a point $p \in
    P$: It is being inserted into the point-set when a node $v$ in the
    HST is being ``split'' at phase $i$ (since $p$ is the
    representative point for one of the children of $v$). Let $p$ and $q$ be
    the two representative points of the two children of $v$. We
    charge $v$ for any work done with $p$ and $q$ for the next $L=10
    \log n $ phases. Consider any work done on $p$ before it
    undergoes another split event. If $p$ is at most $L$ phases
    away from the split event of $v$, the vertex $v$ pays for it.
    
    Otherwise, consider $p$ at $>L$ phases away from its latest split
    event that happened at $v$.  Let $\rcurr$ be the current
    clustering radius, and observe that $p$ represents a set of points
    which has a diameter $\leq \rcurr / n^{2}$ and $\rcurr \leq
    \Delta_v/n^{10}$. In particular, this implies that $ P \cap
    \Ball(p, \rcurr \cdot n^{2}) \subset P \cap \Ball(p,
    \Delta_v/n^{4}) \subset P \cap \Ball(p, \rcurr / n^{2})$.  Namely,
    all the points that $p$ represents are very far from the rest of
    the points of $P$, in terms of $\rcurr$.  In particular, it can
    not be that the cluster that $p$ represents is in any updated
    friends list in the current stage.  (It can be in a friends
    list that was not updated lately, since we use lazy evaluation.
    However, when this friends list will be used, it will be updated
    and $p$ will disappear from it. Note that the work put on updating
    the friends lists is $\lambda^{O(1)} n$ overall, see
    \secref{Gonzalez:low:spread}.)  Thus $p$ does not require any work
    from the algorithm till it undergoes another split event.
    
    Thus, every node in the HST is charged with 
    $\lambda^{O(1)}\log n$ work. It
    follows, that the overall running time of the algorithm 
    $\lambda^{O(1)}n \log n$.
\end{proof}

%----------------------------------------------------------------

\subsection{Constructing the \Nettree{}}
\seclab{net:tree}

In this section we conclude the description of the algorithm for constructing
the \nettree{}, and prove \theoref{net:tree}.

The construction of the \nettree{} $T$ is done by adding points of $P$
according to the \NetPermutAlg{}'s permutation.  As mentioned before,
the construction algorithm and the resulting tree is similar to the
data-structure of Clarkson \cite{c-nnsms-02} (our analysis and the
guaranteed performance are new, however).  The tree constructed for
$p_1,\ldots,p_k$ is denoted by $T^{(k)}$, and $T=T^{(n)}$.  We obtain
$T^{(k)}$ from $T^{(k-1)}$ as follows.

During the construction, we maintain for every vertex $u\in T^{(k)}$ a
set of vertices $\Rel{u}$, which are the vertices close by.  Namely,
the set $\Rel{u}$ would be in fact the set
\[ 
\dRel{u}=\brc{ v\in T^{(k)} \sep{ \mlevel{v}\leq \mlevel{u}<
      \mlevel{\parent(v)}, \text{ and } d_\MTR(\rep_u,\rep_v) \leq
      13\cdot \tbase^{\mlevel{u}}}},
\] 
where $\tbase$ is the packing constant associated with the \Nettree{},
see \defref{net:tree}.  (Since we compute $\Rel{u}$ indirectly, the
fact that $\Rel{u} = \dRel{u}$ requires a formal proof, see
\lemref{correctness} (v).) The set $\Rel{u}$ is of size
$\lambda^{O(1)}$ throughout the algorithm's execution.

We denote by $\br_i=\min\brc{r_j \sep{ 1\leq j \leq i} }$.

\paragraph{The Algorithm.}
The $k$th point in the permutation, $p_k$, will be added as a leaf to
the tree $T^{(k-1)}$ to form the tree $T^{(k)}$. As such we fix
$\mlevel{p_k}=-\infty$, and $\rep_{p_k}=p_k$.  Let
$l=\ceil{\log_\tbase \br_{k-1}}$.

Let $h$ be the largest index such that $\log_\tbase \br_{h-1} > l$
(i.e., $p_h$ is the last added center in the previous phase).  Let
$q\in\brc{p_1,\ldots,p_h}$ the closest point to $p_k$ among
$\brc{p_1,\ldots, p_h}$; namely, $q$ is the nearest neighbor to $p_k$
in all the centers present in the previous phase.  Identifying $q$
with the unique leaf of $T^{(k-1)}$ whose representative is $q$, let
$u=\parent(q)$. We obtain $T^{(k)}$ as follows.
\begin{enumerate}[(a)]
    \item \caselab{diff:level1} If $\mlevel{u}>l$, then we make a new
    vertex $v$, set $\mlevel{v}=l$ and $\rep_v=q$. We then connect $q$
    and $p_k$ as children of $v$, and make $v$ a child of $u$.
    
    \item \caselab{same:level1} Otherwise, connect $p_k$ as another
    child of $u$.
\end{enumerate}

\noindent{}\textbf{Finding $q$}.
Let $c_{p_k}$ be the closest point among $\brc{p_{1},\ldots p_{k-1}}$
to $p_k$ (this information is computed by \NetPermutAlg{}, see
\secref{Gonzalez:low:spread} for details).  Denote $\wu =
\parent(c_{p_k})$.  We consider two cases:
\begin{enumerate}[(1)]
    \item \caselab{diff:level} If $\mlevel{\wu}>l$, then
    $q=\wu$, see \lemref{correctness} \caseref{find:q} for a
    proof.
    
    \item \caselab{same:level} Otherwise, $\mlevel{\wu}=l$.  In this
    case, $q$ must be in the set $\brc{\rep_w \sep{ w\in\Rel{\wu}}}$, see
    \lemref{correctness} \caseref{find:q} for a proof. So, we just
    pick $q$ to be the nearest neighbor to $p_k$ in $\brc{\rep_w\sep{
          w\in\Rel{\wu}}}$.
\end{enumerate}

\noindent{}\textbf{Updating $\Rel{\cdot}$.}
For each new vertex added $x$ we do the following: Let $y=\parent(x)$.
For each $z \in \Rel{y}$, and for each child $z'$ of $z$ we
traverse \emph{part} of the tree rooted at $z'$ in the following way:
When visiting a vertex $u$, we check whether $u$ should be added to
$\Rel{x}$ and whether $x$ should be added to $\Rel{u}$ according to
the $\dRel{\cdot}$ definition, and update $\Rel{x}$ and $\Rel{u}$
accordingly.  If $x$ has been added to $\Rel{u}$ then we continue by
traversing the children of $u$.  Otherwise, we skip them.

Note, that this might require scanning a large fraction of the
\nettree{}, as $x$ might appear in a large number of $\Rel{}$ lists. 

\begin{lemma}
    For any $k \in [1,\ldots,n]$, the tree $T^{(k)}$ has the following
    properties.
    \begin{enumerate}[(i)]
        \item \caselab{find:q} 
        The part of the algorithm that finds $q$, indeed finds it.
        
        \item \caselab{bounded:diam} If $v$ is a child of $u$, then
        $d_\MTR(\rep_u,\rep_v)\leq 2 \cdot \tbase^{\mlevel{u}}$.
        
        \item \caselab{separation} For every $t\in \Re$, every pair of
        points in $\NetC(t)$ is at least $\tbase^{t-1}$ far apart.

        \item \caselab{T:k:net:tree} $T^{(k)}$ is a \nettree{} of
        $\brc{p_1,\ldots,p_k}$.
        
        \item \caselab{Rel:dRel} For any $u\in T$, $\Rel{u}=
        \dRel{u}$.
    \end{enumerate}

    \lemlab{correctness}
\end{lemma}

Since the proof of \lemref{correctness} is tedious, we defer it to
\apndref{net:tree:correct}.  We next analyze the running time.
\begin{lemma}
    Given the (approximate) greedy permutation $\langle p_1,\ldots,
    p_n \rangle$ with their ``current'' cluster's center $\langle
    c_{p_2},\ldots,c_{p_n}\rangle$, the algorithm for constructing the
    \nettree{} runs in $\lambda^{O(1)} n$ time.
\end{lemma}
\begin{proof}
    By the definition of $\Rel{\cdot}$, the size of each
    such list is at most $\lambda^{O(1)}$. Assuming the tree is
    implemented reasonably (with pointers from a vertex to its
    children and parent), the part of constructing the tree clearly
    takes $O(\lambda^{O(1)})$ time per a new point.
    
    Next we estimate the time to construct $\Rel{\cdot}$.  For each
    vertex inserted $x$, we first charge $\lambda^{O(1)}$ visits for
    visiting the children of $\Rel{\parent(x)}$. All the other visits
    are charged to the parent of the visited vertex.  Each vertex
    has at most $\lambda^{O(1)}$ children, and its children are
    visited only if a new entry was inserted to its $\Rel{}$. As the
    total size of the $\Rel{\cdot}$ lists is at most $\lambda^{O(1)}
    n$, we have just bounded the number of visits of vertices during
    the update process of $\Rel{\cdot}$ to $\lambda^{O(1)} n$. Thus
    the time spent is $\lambda^{O(1)} n$.
\end{proof}

%---------------------------------------------------------------------
%---------------------------------------------------------------------

\subsection{Augmenting the \Nettree{}}
\seclab{aug}

In order to efficiently search on the \nettree{}, we will need the
following three auxiliary data structures.

The first one allows, given a vertex $v$ of level $l$, to find all the the
vertices of ``roughly the same level'' which are nearby; i.e., whose
representative is at distance at most $O(\tbase^l)$ from the representative
of $v$.  More accurately, we need a fast access to $\dRel{v}$, as
defined in \secref{net:tree}. We have seen in that section how to
construct it in near linear time such that the whole list can be
accessed in $O(\lambda^4)$ time.

The second data-structure enables the following seek operation: Given
a leaf $x$, and a level $l$, find the ancestor $y$ of $x$ such that
$\mlevel{\parent(y)}> l \geq \mlevel{y}$.  Bender and Farach-Colton
\cite{bf-laps-04} present a data-structure $\CD$ that can be
constructed in linear time over a tree $T$, such that given a node
$x$, and depth $d$, it outputs the ancestor of $x$ at depth $d$ at
$x$. This takes constant time per query. Thus, performing the seek
operation just requires performing a binary search using $\CD$ over the
\nettree{}, and this takes $O(\log n)$ time.

Our third data-structure supports a restricted version of the above
seek operation: Given a leaf $x$, an ancestor $z$ of $x$, and given a
level $l$: If $l \notin [\mlevel{z}- c \log n, \mlevel{z}]$ return
``don't know''.  Otherwise, return an ancestor $y$ of $x$ satisfying
$\mlevel{\parent(y)} > l \geq \mlevel{y}$ (here $c>0$ is an absolute
constant).  The data structure has $O(n)$ space, $O(n \log n)$
preprocessing time, and the queries can be answered in \emph{constant
   time}.

As a first step, observe that if for every internal vertex $z$ and a
descendant leaf $x$ we add vertices to the tree so as to fill all
levels between $\mlevel{z}$ and $\mlevel{x}- c \log n$ on the path
between $z$ and $x$, then queries to $l$ level ancestor, $l \in
[\mlevel{x}- c \log n, \mlevel{z}]$ can be answered by using the data
structure $\CD$ as above to find an ancestor of $x$ at depth
$\depth(z)-(\mlevel{z}-l)$. This construction, however, may blow up
the number of vertices in the \nettree{} (and hence the space) by a
$\log n$ factor.

To obtain linear space we do the following: In the preprocessing step
we enumerate all the possible patterns of existence/nonexistence of
vertices in $0.5 \log_2 n$ consecutive levels. For each given pattern,
and each given level in the pattern we write the number of actual
vertices above this level.  Preparing this enumeration takes
only $O(\sqrt{n} \log n)$ time.  Now, for each vertex $u$ of the
\nettree{}, we hold $2c$ pointers to such patterns that together map
the vertices in the $c \log n$ level below $v$ on the path to $u$,
where $v$ is an ancestor of $u$ at \emph{depth} $\depth(u)-c \log n$,
if such $v$ exists (note that $v$ is $c\log n$ edges above $u$ in the
\nettree{}, but $u$ holds the pattern of only the first $c \log n$
\emph{levels} below $v$).  This data structure can be clearly computed
in $O(n \log n)$ time using top-down dynamic programming on the
\nettree{}.

Given a query (with $x$, $z$ and $l$ as above), we do as follows: Let
$u$ be an ancestor of $x$ at depth $\max\sbrc{\depth(z)+c\log n,
   \depth(x)}$. Vertex $u$ can be accessed in $O(1)$ time using the
data-structure $\CD$. Using the patterns pointed by $u$ we can find
the depth of the relevant vertex whose level is just below $l$ in
$O(1)$ time, and now using $\CD$ again we can access this vertex in
constant time.

%-----------------------------------------------------
%-----------------------------------------------------
%-----------------------------------------------------

\section{Approximate Nearest-Neighbor Search}
\seclab{nn:n:l:s}

In the following, ANN stands for approximate nearest neighbor.  In
this section, we present an approximate nearest neighbor (ANN) scheme,
that for a given set of points $P$, preprocess it in near linear time,
and produce a linear space data-structure which answers queries of the
form ``given point $q$, find $p\in P$, such that $d(q,p)\leq (1+\eps)
d(q,P)$'' in logarithmic time.  See \secref{intro} for more details.

In \secref{ann:low-spread}, we present a variant of Krauthgamer and
Lee \cite{kl-nnsap-04} net navigation ANN algorithm that works on the
\nettree{}. This algorithm allows to boost an $A$-ANN solution to
$(1+\eps)$-ANN solution in $O(\log n+ \log (A/\eps))$ query time.  In
\secref{lq-rst} we present a fast construction of a variant of the
ring separator tree \cite{im-anntr-98,kl-bbcnn-04}, which support fast
$2n$-ANN queries.  We conclude in \secref{ann:general} with the
general scheme which is a combination of the previous two.

\subsection{The Low Spread Case}
\seclab{ann:low-spread}

\begin{lemma}
    Given a \nettree{} $T$ of $P$, a query point $q\in \MTR$, and
    vertex $u\in T$ at level $l=\mlevel{u}$ such that
    $d_\MTR(\rep_u,q)\leq 5 \cdot \tbase^{l}$ or $\widehat{p}\in P_u$, 
    where $\widehat{p}$ is the nearest neighbor to $q$ in $P$.
    Then there is an algorithm that traverse $T$ form $u$ downward, 
    such that for any $t\in \mathbb{N}$,
    after $t+4$ steps, the
    algorithm reaches a vertex $s$ 
    for which $\rep_s$ is a $(1+\tbase^{l-f-t})$-ANN,
    where $f= \log_\tau d_\MTR(\widehat{p},q)$.
    The running time of this search is $\lambda^{O(1)} \min\brc{t,l-f}
    + \lambda^{O(\max\sbrc{t - (l-f),0 } )}$.
    
    \lemlab{nn:bounded}
\end{lemma}
\begin{proof}
    The query algorithm works as follows.  It constructs sets $A_i$ of
    vertices in $T$ with the following properties:
    \begin{enumerate}
        \item For each $v\in A_i$, $\mlevel{\parent(v)}> i \geq
        \mlevel{v}$.
        
        \item $\widehat{p} \in \cup_{v\in A_i} P_v \subset
        \Ball(q,d_\MTR(q,\widehat{p})+(13+\tfrac{2\tbase}{\tbase-1})
        \cdot \tbase^{i})$.
    \end{enumerate}
    
    The algorithm starts by setting $A_l=\Rel{u}$.  If $\widehat{p}\in
    P_u$ then $A_l$ clearly satisfies the two properties above.  If
    $d_\MTR(\rep_u,q)\leq 5\cdot \tbase^l$, then
    $d_\MTR(\rep_u,\widehat{p})\leq 10 \tbase^l$.  Suppose for the
    sake of contradiction that $\widehat{p} \notin \cup_{v\in A_l}
    P_v$, then $\exists v'$ such that $\mlevel{v'}\leq l$,
    $d_\MTR(\rep_u,\rep_{v'})>13 \tbase^l$, and $\widehat{p}\in
    P_{v'}$. But then from the covering property
    $d_\MTR(\rep_{v'},\widehat{p}) \leq \tfrac{2\tbase}{\tbase-1}
    \tbase^l$ which means that $d_\MTR(\rep_u,\widehat{p})>(13 -
    \tfrac{2\tbase}{\tbase-1} ) \tbase^l > 10 \tbase^l$, a
    contradiction.
    \begin{comment}
    The second property is satisfied, since
   the packing property of the \nettree{} implies that
    \[
    P\cap \Ball(q,d_\MTR(q,\rep_u)) \subset
    P\cap \Ball(\rep_u,2d_\MTR(q,\rep_u))
    \subset P \cap \Ball(q,\tfrac{1}{4}\cdot \tbase^{\mlevel{\parent(u)}-1})
    P_u
    \]

    which clearly
    satisfies the properties above.
    \end{comment}
    
    The set $A_{i-1}$ is constructed from $A_i$ as follows: Let $v\in
    A_i$ be the closest vertex in $A_i$ to $q$, i.e.,
    $d_\MTR(\rep_v,q)=\min _{w\in A_i} d_\MTR(\rep_w,q)$.  Let $B$ the
    set obtained from $A_i$ by replacing every vertex of level $i$ with its
    children.  The set $A_{i-1}$ is obtained from $B$ by throwing out 
    any vertex $w$ for which $d_\MTR(q,\rep_w)> d_\MTR(q,\rep_v)+
    \tfrac{2\tbase}{\tbase-1}\cdot \tbase^{i-1}$.  
    It is easily checked that $A_{i-1}$ has the required
    properties.
    
    The running time is clearly dominated by $\lambda^{O(1)}$ times
    the sum of the $A_i$'s sizes.  For $i> f$, $d_\MTR(q,\rep_v)$ is
    at most $\tfrac{2\tbase}{\tbase-2}\cdot \tbase^{i}$, 
    and therefore $|A_i|\leq \lambda^{O(1)}$. For
    $i\leq f$, we have only a weak bound of $|A_i|\leq
    \lambda^{O(f-i)}$.  Thus the running time of the algorithm for $t$
    steps follows. Notice that any point in $A_{l-i}$ is 
    $(1+\tbase^{l-f-i+4})$-ANN.
\end{proof}

\smallskip

For a set $P$ with spread $\Spread$, by applying the algorithm of
\lemref{nn:bounded} with $u$ the root of $T$, and $t=\ceil{\log_\tbase
   (\Spread/\eps ) -f}$, \lemref{nn:bounded} gives a
$(1+\eps)$-approximate nearest neighbor scheme with $O(n \log n)$
expected construction time and $O(\log \Spread + \eps^{-O(\dim)})$
query time.  (Note that the algorithm does not need to know $t$ (and
thus $f$) in advance, it can estimate the current approximation by
comparing $d_\MTR(q,\rep_v)$ to $\tbase^i$.)  This gives an
alternative to the data-structure of Krauthgamer and Lee
\cite{kl-nnsap-04}, with a slightly faster construction time. Their
construction time is $O( n \log \Spread \log \log \Spread)$ if one
uses the insertion operation for their data-structure (note that in
the constant doubling dimension setting, $\log n=O(\log \Spread)$).
In fact, in this case, the $\Rel{}$ data-structure is not needed since
$\Rel{\Root}=\brc{\Root}$. Therefore the storage for
this ANN scheme is $O(n)$, with no dependency on the dimension.  A
similar construction was obtained independently in \cite{bkl-ctnn-04}.
However, their construction time is $O(n^2)$.

\subsection{Low Quality Ring Separator Tree}
\seclab{lq-rst}

\begin{lemma}
    One can construct a data-structure which supports $2n$-ANN queries
    in $2^{O(\dim)}$ $\log n$ time. The construction time is
    $2^{O(\dim)} n \log n $, and the data-structure uses $2^{O(\dim)}
    n$ space.
\end{lemma}
\begin{proof}
    The data structure is a binary search tree $S$, in which each
    vertex of the tree $v$ is associated with a point $p_v\in P$
    and radius $r_v$.  We are guaranteed that $n/2\lambda^{3} \leq
    |\Ball(p_v,r_v)| \leq (1-1/2\lambda^{3})n$, and that
    $\pth{\Ball(p_v,(1+1/2n)r_v) \setminus \Ball(p_v,(1-1/2n)r_v)} \cap P 
    = \emptyset$.
    The left subtree is
    recursively constructed on the set $P \cap \Ball(p_v,r_v)$, and
    the right subtree is recursively constructed on $P\setminus
    \Ball(p_v,r_v)$.  The depth of $S$ is clearly at most $O(\lambda^3
    \log n)$.
    
    The construction of $S$ is similar to the construction of the
    low-quality spanner (\secref{HST}) and uses \lemref{small:ball} as
    follows.  Apply \lemref{small:ball} to find $p\in P$ and $r$ such
    that $|\Ball(p,r)|\geq n/(2\lambda^3)$, whereas $|\Ball(p,2r)|\leq
    n/2$. From the pigeon-hole principle, there exists
    $r'\in[(1+1/2n)r,2r-r/2n)$ for which $\Ball(p,(1+1/2n)r')\setminus
    \Ball(p,(1-1/2n)r')= \emptyset$. We now make a root $v$ for the
    ring separator tree, set $p_v=p$, and $r_v=r'$, and recurse on
    $\Ball(p_v,r_v)$ as the left subtree, and $P\setminus
    \Ball(p_v,r_v)$ as the right subtree.
    The construction time $T(n)$ obeys the recursive formula
    $T(n)=T(n_1)+T(n_2)+O(n)$, where $n_1+n_2=n$, 
    $n/2\lambda^3 \leq n_1\leq n/2$.
    
    Once we have this data-structure, $2n$-ANN can be found in
    $O(\lambda^3 \log n)$ time as follows.
    Let the root of the ring separator tree be $u$.  Given a query
    point $q$, check its distance to $p_u$.  If $d_\MTR(q,p_u)\leq r_u$
    then recurse on the left subtree.  Otherwise, recurse on the right
    subtree.  At the end, return the nearest point to $q$ among $p_v$,
    where $v$ is on the path traversed by the algorithm.
    
    The running time of this procedure is clearly dominated by the
    height of the tree which is $O(\lambda^3 \log n)$.
    
    To see that this is indeed $2n$-ANN, let $a$ be the vertical path
    in the tree traversed by the algorithm, and let $b$ be the
    vertical path in the tree connecting the root to the nearest
    neighbor of $q$ in $P$. Let $v$ be the lowest common vertex of $a$
    and $b$. Suppose that $a$ continued on the left subtree of $v$
    while $b$ continued on the right subtree. In this case the
    distance from $q$ to the nearest neighbor is at least $r_v/2n$,
    while $d_\MTR(p_v,q)\leq r_v$. Thus $p_v$ is $2n$-ANN.
    
    If $a$ continued on the right subtree of $v$ while $b$ continued
    on the left subtree of $v$, then The distance from the nearest
    neighbor is at least $r_v/2n+ (d_\MTR(p_v,q)-r_v)$, while $p_v$ is
    at distance $d_\MTR(p_v,q)$.  The ratio between this two
    quantities is clearly at most $2n$.
\end{proof}

\begin{remark}
    As is pointed out in \cite{im-anntr-98,kl-bbcnn-04}, it is
    possible to duplicate points in the ring for the two subtrees.
    Hence we can actually partition the $\Ball(p,2r)\setminus
    \Ball(p,r)$ into $t\leq n$ sub rings, and choose to duplicate a
    ``light'' ring.  When $t=1$, we obtain the Ring Separator Tree
    from \cite{kl-bbcnn-04}, that supports $O(1)$-ANN queries, but
    requires $n^{2^{O(\dim)}}$ storage.  For general $t\leq n $ we
    obtain a data structure that supports $O(t)$-ANN queries, and by
    choosing the right ring to duplicate, consumes only $n^{(3\log 2
       \lambda)^{1/t}}$ storage.  To see this, we set $\beta={(3\log 2
       \lambda)^{1/t}}$ and prove by induction on $n$ that it is
    possible to find a ring such that the number of leaves in the tree
    is at most $n^\beta$.  Denote by $\eta_i =|\Ball(p,(1+i/t)r)|/n$.
    Note that $(2\lambda)^{-3}\leq \eta_0\leq\eta_1\leq \cdots \eta_t
    \leq n/2$, and therefore there exists $i\leq t$ for which
    $\eta_{i-1} \geq \eta_i^\beta$, otherwise $(2\lambda)^{-3} <
    \eta_0^{\beta^t} \leq (1/2)^{\beta^t}$ which is a contradiction.
    Thus by duplicating the $i$th ring, and by applying the inductive
    hypothesis on the number of leaves in the subtrees, the resulting
    tree will have at most $(\eta_i n)^\beta+
    ((1-\eta_{i-1})n)^\beta \leq (\eta_{i-1}+(1-\eta_{i-1})) n^\beta$
    leaves.
    
    Thus, setting $t=O(\log\log \lambda \cdot \log n)$, we obtain a
    linear space ring separator tree that supports $O(t)$-ANN queries in
    $O( \log n )$ time.
\end{remark}

\subsection{ANN algorithm for arbitrary spread}
\seclab{ann:general}

The algorithm for arbitrary spread is now pretty clear. During the
preprocessing we construct the augmented \nettree{} from
\secref{nets}.  We also construct the low quality ring separator tree.
The construction time is $2^{O(\dim)} n \log n$, and the space used is
$2^{O(\dim)} n$.

Given a query point $q\in\MTR$, and the approximation parameter $\eps>0$,
the query algorithm consists of three steps:
\begin{enumerate}
    \item First, find $2n$-ANN $p_1$ using the low quality ring
    separator tree of \secref{lq-rst}.
    
    \item Next find a vertex $u$ in the \nettree{} which is an
    ancestor for $p_1$ and that satisfies
    \[
    \mlevel{\parent(u)}-1\geq
    \ceil{\log_\tbase (16\cdot d_\MTR(p_1,q))}\geq \mlevel{u}. 
    \]
    Hence
    \[
    d_\MTR(\rep_u,q)\leq d_\MTR(\rep_u,p_1) + d_\MTR(p_1,q) \leq 2.5
    \cdot \tbase^{\mlevel{u}} + \tfrac{1}{16}
    \tbase^{\mlevel{\parent(u)}-1}.
    \]
    
    \item We now split the analysis into two cases.
    \begin{enumerate}
        \item If $2.5 \cdot \tbase^{\mlevel{u}} \geq \tfrac{1}{16}
        \tbase^{\mlevel{\parent(u)}-1}$, then clearly
        $d_\MTR(\rep_u,q) \leq 5 \tbase^{\mlevel{u}}$, and thus $u$
        satisfies the conditions of \lemref{nn:bounded}.
        
        \item If on the other hand $2.5\cdot \tbase^{\mlevel{u}} <
        \tfrac{1}{16} \tbase^{\mlevel{\parent(u)}-1}$, then the
        packing property of the \nettree{} implies that
        \[
        P\cap \Ball(q,d_\MTR(q,\rep_u)) \subset P\cap
        \Ball(\rep_u,2d_\MTR(q,\rep_u)) \subset P \cap
        \Ball(\rep_u,\tfrac{1}{4}\cdot \tbase^{\mlevel{\parent(u)}-1})
        \subset P_u ,
        \]
        and therefore $\widehat{p}\in P_u$. Thus, in this case $u$
        also satisfies the conditions of \lemref{nn:bounded}.
    \end{enumerate}
    \item Set $l=\mlevel{u}$. Using the notation of
    \lemref{nn:bounded}, the fact that $p_1$ is a $2n$-ANN, implies
    that $f\geq l-(1+\log n)$, thus by setting the number of steps to
    $t=\ceil{\log (n/\eps)}$, and applying the algorithm of
    \lemref{nn:bounded}, we obtain $(1+\eps)$-ANN.
\end{enumerate}
The running time of the query is
\[
\lambda^{O(1)} \log n +O(\log n)+\lambda^{O(1)}\log n+\eps^{-O(\dim)}
\leq \lambda^{O(1)} \log n+ \eps^{-O(\dim)} .
\]

We summarize:
\begin{theorem}
    Given a set $P$ of $n$ points of bounded doubling dimension $\dim$
    in a metric space $\MTR$, One can construct a data-structure for
    answering approximate nearest neighbor queries (where the quality
    parameter $\eps$ is provided together with the query).  The query
    time is $2^{O(\dim)} \log n + \eps^{-O(\dim)} $, the 
    expected preprocessing
    time is $2^{O(\dim)} n \log n $, and the space used is
    $2^{O(\dim)} n $.  

    \theolab{nn:general}
\end{theorem}

\theoref{nn:general} compares quite favorably with the result of
Krauthgamer and Lee \cite{kl-bbcnn-04}, which solves the same problem
with the same (tight) query time but using $O( 2^{O(\dim)} n^2$
$\polylog(n))$ space.

%------------------------------------------------------------------
%------------------------------------------------------------------
%------------------------------------------------------------------

\section{Fast construction of \WSPD{} and Spanners}
\seclab{WSPD}

Let $P$ be an $n$-point subset of a metric space $\MTR$ with doubling
dimension $\dim$, and $1/4>\eps >0$ a parameter.  Denote by $A \otimes
B$ the set $\brc{\brc{x,y} \sepS{ x\in A,\, y\in B}}$.  A
\emph{well-separated pair decomposition} (\WSPD{}) with parameter
$\eps^{-1}$ of $P$ is a set of pairs
$\brc{\brc{A_1,B_1},\ldots,\brc{A_s,B_s}}$, such that
\begin{enumerate}
    \item $A_i,B_i\subset P$ for every $i$.

    \item $A_i \cap B_i = \emptyset$ for every $i$.

    \item $\cup_{i=1}^s A_i \otimes B_i = P \otimes P$.
    
    \item $\DstM(A_i,B_i) \geq \eps^{-1} \cdot \max \brc{ \diam(A_i),
    \diam(B_i) }$
\end{enumerate}

The notion of \WSPD{} was defined by Callahan and Kosaraju
\cite{ck-dmpsa-95} for Euclidean spaces.  Talwar \cite{t-beald-04}
have shown that this notion transfer to constant doubling metrics. In
particular, he proves that any $n$-point metric with doubling dimension
$\dim$ admits \WSPD{} in which the number of pairs is
$n\eps^{-O(\dim)}\log \Spread$.  We improve this result.

\begin{lemma} 
    For $1 \geq \eps>0$, one can construct a $\eps^{-1}$-\WSPD{} of size
    $n \eps^{-O(\dim)}$, and the expected construction time is
    $2^{O(\dim)}n \log n + n \eps^{-O(\dim)}$.
    
    Furthermore, the pairs of the \WSPD{} correspond to $(P_u,P_v)$,
    where $u,v$ are vertices of a \nettree{} of $P$, and for any pair
    $(P_u,P_v)$ in \WSPD{}, $\diam(P_u),\diam(P_v) \leq \eps
    d_P(\rep_u,\rep_v)$.

    \lemlab{WSPD}
\end{lemma} 
\begin{proof} 
    We compute the \nettree{} $T$ using \theoref{net:tree}.  For
    concreteness of the \WSPD{}, assume also that some weak linear
    order $\preceq$ is defined on the vertices of $T$.  The \WSPD{} is
    constructed by calling to $\WSPDProc(u_0,u_0)$, where $u_0$ is the
    root of the \nettree{} $T$, and $\WSPDProc(u,v)$ is defined
    recursively as follows.
    \begin{center}
        \fbox{
           \begin{minipage}{6cm}
               \begin{tabbing}
                   \ \ \ \= \ \ \ \= \ \ \ \= \kill
                   $\WSPDProc(u,v)$\+\\
                   Assume $\mlevel{u}> \mlevel{v}$ or
                   ( $\mlevel{u}= \mlevel{v}$ and $u \preceq v$) \\
                   \> \> (otherwise exchange $u\leftrightarrow v$). \\
                   If $8\tfrac{2\tbase}{\tbase-1}\cdot
                   \tbase^{\mlevel{u}}\leq {\eps} 
                   \cdot \DstM( \rep_u, \rep_v) $ then \\
                   \> \Return $\brc{ \MakeSBig \brc{u,v} \, }$\\ 
                   \Else \+\\
                   Denote by $u_1,\ldots,u_r$ the children of $u$\\
                   \Return $\bigcup_{i=1}^r \WSPDProc(u_i,v)$. 
               \end{tabbing}
           \end{minipage}}
    \end{center}    

    For any node $u \in T$ we have $\diam(P_u) \leq 2
    \tfrac{2\tau}{\tau-1} \cdot \tbase^{\mlevel{u}}$ (see
    \defref{net:tree}).  In particular, for every output pair
    $\brc{u,v}$ it holds
    \begin{eqnarray*}
        \max\sbrc{\diam(P_u),\diam(P_v)} &\leq& 2 \tfrac{2\tbase}{\tbase-1}
        \cdot \max\sbrc{\tbase^{\mlevel{u}}, \tbase^{\mlevel{v}}} \leq
        \tfrac{\eps}{4} d_P(\rep_u,\rep_v) \\
        &\leq& \tfrac{\eps}{4} ( d_P(P_u,P_v)+ \diam(P_u)+\diam(P_v)),
    \end{eqnarray*}
    and so $\max\sbrc{\diam(P_u),\diam(P_v)} \le
    \frac{\eps}{4(1-\eps/2)} d_P(P_u,P_v) \leq \eps d_P(P_u,P_v)$,
    since $\eps \leq 1$.  Similarly, for any $x \in P_u$ and $y \in
    P_v$, we have
    \[
    d_P( \rep_u, \rep_v) \leq d_P(x,y) + \diam(P_u) + \diam(P_v) 
    \leq ( 1+\eps) d_P(x,y).
    \]

    One can verify that every pair of points is covered by a pair of
    subsets $\brc{P_u,P_v}$ output by the \WSPDProc{} algorithm.
    
    We are left to argue about the size of the output (the running
    time is clearly linear in the output size). Let $\brc{u,v}$ be an
    output pair and assume that the call to $\WSPDProc(u,v)$ was
    issued by $\WSPDProc(u,\parent(v))$. We charge this call to
    $\parent(v)$, and we will prove that each vertex is charged at
    most $\eps^{-O(\dim)}$ times.
    
    Fix $v'\in T$. It is charged by pairs of the form $\brc{u,v}$ in
    which $\parent(v)=v'$, and which were issued inside
    $\WSPDProc(u,v')$.  This implies that $\mlevel{\parent(u)} \geq
    \mlevel{v'} \geq \mlevel{u}$.
    
    Since the pair $(u,v')$ was not generated by \WSPDProc{}, it must
    be that conclude that $d_P(\rep_{v'},\rep_u) \leq
    8\frac{2\tbase}{\tbase-1} \cdot \tbase^{\mlevel{v'}}/\eps$.  The
    set
    \[
    U=\brc{w \sep{ \mlevel{\parent(w)} \geq \mlevel{v'} \geq
          \mlevel{w}} \text{ and } d_P(\rep_{v'},\rep_w) \leq
       8\frac{2\tbase}{\eps(\tbase-1)} \cdot \tbase^{\mlevel{v'}} }
    \]
    contains $u$, and $U$ is a subset of $\NetC({\mlevel{v'}})$. By
    \propref{net:tree}, for every $u_1,u_2\in U$, if $u_1\neq u_2$
    then $d_P(P_{u_1},P_{u_2}) \geq \tbase^{\mlevel{v'}-1}/4$.  By the
    doubling property, we have $|U|\leq \eps^{-O(\dim)}$.  We 
    therefore infer that $v'$ can only be charged by pairs in
    $U\times C_{v'}$,
    where $C_{v'}$ is the set of children of $v'$. 
    We conclude that $v'$ might be charged at most
    $|U|\cdot |C_{v'}| \leq (2/\eps)^{O(\dim)} = \eps^{-O(\dim)}$ times. 
     Thus, the total
    number of pairs generated by the algorithm is $n\eps^{-O(\dim)}$.
\end{proof}

\subsection{Spanners}

\begin{defn}
A $t$-spanner of a finite metric space $P$ is a weighted
graph $G$ whose vertices are the points of $P$, and for any $x,y\in P$,
\[ 
d_P(x,y)\leq d_G(x,y) \leq t \cdot d_P(x,y), 
\]
where $d_G$ the metric of the shortest path on $G$.
\end{defn}

\begin{theorem} \theolab{spanner}
    Given an $n$-point metric $P$ with doubling dimension $\dim$, and
    parameter $1 \geq \eps >0$, one can compute a $(1+\eps)$-spanner
    of $P$ with $n\eps^{-O(\dim)}$ edges, in $2^{O(\dim)}n \log n + n
    \eps^{-O(\dim)}$ expected time.
\end{theorem}
\begin{proof}
    Let $c\geq 16$ be an arbitrary constant, and set $\delta=\eps/c$.
    Compute a $\delta^{-1}$-\WSPD{} decomposition using the algorithm of
    the previous section.  For every pair $\brc{u,v}\in \WSPD$, add an
    edge between $\brc{\rep_u,\rep_v}$ with weight
    $d_P(\rep_u,\rep_v)$.  Let $G$ be the resulting graph, clearly,
    the resulting shortest path metric $d_G$ dominates the metric
    $d_P$.
    
    The upper bound on the stretch is proved by induction on the length
    of pairs in the WSPD.  Fix a pair $x,y\in P$, by our induction
    hypothesis, we have for every pair $z,w \in P$ such that
    $d_P(z,w) < d_P(x,y)$, it holds $\Dist_G(z,w) \leq
    (1+c\delta)\Dist_P(z,w)$.
    
    The pair $x,y$ must appear in some pair $\sbrc{u,v} \in \WSPD$,
    where $x\in P_u$, and $y\in P_v$.  Thus $d_P(\rep_u,\rep_v)\leq
    (1+2\delta)d_P(x,y)$ and $d_P( x, \rep_u),
    \DstM(y,\rep_v) \leq \delta \DstM(\rep_u, \rep_v)$, by
    \lemref{WSPD}. By the inductive hypothesis
    \begin{eqnarray*}
        d_G(x,y) &\leq& d_G(x,\rep_u) 
        +  d_G(\rep_u,\rep_v) 
        +  d_G(\rep_v,y)\\
        &\leq&  
        (1+c\delta) d_P(x,\rep_u)+ d_P(\rep_u, \rep_v) 
        + (1+c\delta) d_P(\rep_v,y)\\
        &\leq&  
        2(1+c\delta) \cdot \delta \cdot d_P(\rep_u, \rep_v )
        + d_P(\rep_u, \rep_v) \\
        &\leq&  %(1  + 2\delta + 2c \delta^2 ) d_P(\rep_u,\rep_v )
         % \leq 
        (1  + 2\delta + 2c \delta^2 ) (1+2\delta)d_P( x, y )\\
        % &\leq& (1  + 4\delta ) (1+\delta)d_P( x, y )
        % \leq (1  + 5\delta +4\delta^2)d_P( x, y )\\
        &\leq & %(1  + c\delta)d_P( x, y ) \leq 
       (1+\eps)d_P(x,y),
    \end{eqnarray*}
    since $\delta c \leq \eps \leq 1$ and $16 \delta \leq 1$ and $c
    \geq 11$.
\end{proof}

\section{Compact Representation Scheme}
\seclab{CRS}

A \emph{compact representation scheme} (\emph{CRS}) of a finite metric
space $P$ is a ``compact'' data-structure that can answer distance
queries for pairs of points. We measure the performance
of a CRS using four parameters $(\sP,\sS,\sQ,\sAprx \,)$, where $\sP$
is the preprocessing time of the distance matrix, $\sS$ is the space
used by the CRS (in terms of words), $\sQ$ is the query time, and
$\sAprx$ is the approximation factor.

The distance matrix by itself is a $(\sP=O(1),\sS=O(n^2), \sQ=O(1),
\sAprx=1)$-CRS.  The $\eps^{-1}$-WSPD, as well as the $(1+\eps)$-spanner
are representations of $(1+O(\eps))$-approximation of the metric that
consumes only $\eps^{-O(\dim)}n$ space.  However, na\"ively it takes
$\Omega(n)$ time to answer approximate distance queries in these
data-structures.

In this section, we obtain the following theorem.
\begin{theorem}   
    For any $n$ point metric with doubling dimension $\dim$, there
    exist:
    \begin{enumerate}[(a)]    
        \item $(\sP=2^{O(\dim)}n \log^2 n + \eps^{-O(\dim)} n,
        \sS=\eps^{-O(\dim)} n, \sQ=2^{O(\dim)}, \sAprx=1+\eps)$-CRS.
	\label{item:crs_a}
        
        \item $(\sP=2^{O(\dim)}\cdot \poly(n) + \eps^{-O(\dim)} n,
        \sS=\eps^{-O(\dim)} n, \sQ={O(\dim)}, \sAprx=1+\eps)$-CRS.
	\label{item:crs_b}
    \end{enumerate}

    \theolab{CRS}
\end{theorem}

For general $n$-point metrics, Thorup and Zwick \cite{tz-ado-01}
obtained a $(k n^{1+1/k}, kn^{1+1/k}, O(k),$ $2k-1)$-CRS, where $k\in
\NN$ is a prescribed parameter.  The trade-off between the
approximation and the space is essentially tight for general metrics.
Closer in spirit to our setting, Gudmunsson \etal{}
\cite{glns-adogg-02,glns-ador-02} considered metrics that are $t$
approximated by Euclidean distances in $\Re^d$, where both $d$ and $t$
are (possibly large) constants.  They showed that such metrics have
$(O(n \log n), O(n), O(1), 1+\eps)$-CRS (The $O$ notation here hides
constants that depend on $\eps$, $d$ and $t$).  Our scheme strictly
extends%
\footnote{Caveat: They use a weaker model of computation.}  their
result since metrics that are $t$ approximated by a set of points in
the $d$-dimensional Euclidean space has doubling dimension at most
$d\log(2t)$.  We further discuss previous work on special type of CRS,
called \emph{distance labeling}, in \secref{dist-label}.

Our scheme is naturally composed of two parts: In \secref{do:net:tree}
we show how using the \nettree{} it is possible to convert an
$A$-approximate CRS into $(1+\eps)$-approximate CRS in essentially
$O(\log A)$ query time (and even $O(\log \log A)$ query time).  We
then show in \secref{do:Assouad} how to obtain $O(1)$-approximate CRS
using Assouad's embedding.  In \secref{dist-label} we observe that
Assouad's embedding can be used in distance labeling schema.

% note that Assouad's embedding can be used
% to unify two recent results of Talwar~\cite{t-beald-04} 
% and Slivkins~\cite{s-date-05} on distance labeling.

\subsection{Approximation Boosting Lemma}
\seclab{do:net:tree}

Assume we are given a data structure $\CA$, which is
$(\sP,\sS,\sQ,\sAprx)$-CRS of a set ${P}\subset \MTR$, where $\sAprx
\leq 3n^2$.  In this section, we derive a CRS with improved
approximation.  Besides storing the data-structure of $\CA$, we also
need the following data structures:

\begin{enumerate}
    \item The \nettree{} $T$ augmented so that it supports the
    following operations:
    \begin{enumerate}
        \item $O(\log n)$ time access for ancestors of given level as
        defined in \secref{aug}.
        
        \item Constant time access for ancestor of given $x$, when the
        level is at most $6 \log n$ levels below a given ancestor $z$.
        Again, \secref{aug} contains more information.
            
        \item A constant time access for the $\lca{}$ of two vertices
        in $T$ \cite{bf-lpr-00}.
    \end{enumerate} 
        
    \item A $\eps^{-1}$-\WSPD{} $W$ on the \nettree{} $T$, with support for fast
    membership queries.  For each pair we also store the distance
    between their representatives.  Using hashing % \cite{clrs-ia-01},
    membership queries can be answered in constant time.
    
    \item The $(3n^2)$-approximation HST $H$ of \secref{HST}.  The HST
    $H$ should be augmented with the following features:
    \begin{enumerate}
        \item A constant time access to least-common-ancestor queries,
        after a linear time preprocessing \cite{bf-lpr-00}.
            
        \item Each vertex $u$ of $H$ contains pointers to the
        following set of vertices in $T$
        \[
        K_u=\brc{x\in T:\ d_\MTR(\rep_x,\rep_u)\leq 4 \Delta_u
           \text{ and } \mlevel{x}<\log \Delta_u \leq
           \mlevel{\parent(x)} }.
        \]
        Note that $|K_u| \leq \lambda^{O(1)}$, and computing all these
        sets can be accomplished in $\lambda^{O(1)} n \log
        n$-time by finding the level $\ceil{\log \Delta_u}$ ancestor
        $z$ of $\rep_u$ in $T$ in $O(\log n)$ time, and then scanning
        $\Rel{z}$.
    \end{enumerate}
\end{enumerate}    
All these data-structures can be created in 
$2^{O(\dim)}n \log n +\eps^{-O(\dim)}n$ time and $\eps^{-O(\dim)} n$
space.
    
Assuming Query-$\CA(x,y)$ returns a value $\eta$, such that
$d_\MTR(x,y)/ \sAprx\leq \eta \leq d_\MTR(x,y)$, the query
algorithm is:
\begin{center}
    \fbox{
       \begin{minipage}{6cm}
           \begin{tabbing}
              \ \ \ \  \ \ \ \= \ \ \ \ \ \= \ \ \ \  \ \= \kill
               Query-$\CB(x,\ y\in P)$\+\\
               $z \leftarrow \lca{}_H(x,y)$.\\
               $u' \leftarrow$ ancestor of $x$ in $T$ among $K_z$,
               $v' \leftarrow$ ancestor of $y$ in $T$ among $K_z$.\\
               $\eta \leftarrow $ Query-$\CA(x,y)$.\\
               $u_0\leftarrow$ ancestor of $x$ in level $\floor{\log
               (\eps \eta) }$,
               $v_0\leftarrow$ ancestor of $y$ in level $\floor{\log(
               \eps \eta) }$.\\
               $u\leftarrow u_0$, $v\leftarrow v_0$.\\
               \While $\brc{u,v}\notin W$ do \+ \\
               \If $\mlevel{\parent(u)}<\mlevel{\parent(v)}$ or
               ( $\mlevel{\parent(u)}=\mlevel{\parent(v)}$
               and $\parent(v) \preceq \parent(v)$ ) \Then\\
               \> $u \leftarrow \parent(u)$ \\
               \Else\\
               \> $v \leftarrow \parent(v)$.\-\\
               \Return $d_\MTR(\rep_u,\rep_v)$.\- 
           \end{tabbing}
       \end{minipage}}
\end{center}

Implementation details: $u'$ is found by scanning all vertices in
$K_z$ (there are only $\lambda^{O(1)}$ such vertices), and checking
which one of them is an ancestor of $x$ in $T$ (ancestorship can be
checked using the $\lca$ operation on $T$). Note that an ancestor of
$x$ must be contained in $K_z$, since $d_\MTR(\rep_z,x)\leq \Delta_z$,
and thus the ancestor of level immediately below $\log \Delta_z$ must
be in $K_z$.  Similar thing happens with $v'$.  Both $\eta$ and
$\Delta_z$ are $3n^2$ approximation to $d_\MTR(x,y)$ and therefore
$\mlevel{u'}-\mlevel{u_0}\leq 4 \log n+3$, hence $u_0$ can be accessed
in constant time. The same goes to $v_0$.

The following lemma is immediate consequence of the way the WSPD
algorithm works.

\begin{lemma}
    For a pair $\brc{s,t}\in W$ (the $\eps^{-1}$-WSPD), and
    $\mlevel{s}\leq \mlevel{t}$, one of the following conditions must
    be satisfied:
    \begin{enumerate}
        \item $\mlevel{s}\leq \mlevel{t} <\mlevel{\parent(s)}$ and
        $\tfrac{2\tbase}{\tbase-1}\cdot
        \tbase^{\mlevel{\parent(s)}} > \eps \cdot
        d_\MTR(\rep_{\parent(s)}, \rep_t)$, and
        $\tfrac{2\tbase}{\tbase-1} \cdot \tbase^{\mlevel{s}} \leq
        \eps \cdot d_\MTR(\rep_{s}, \rep_t)$.
        
        \item $\mlevel{s}< \mlevel{t}= \mlevel{\parent(s)}$, and
        $\parent(s)\preceq t$, and $\tfrac{2\tbase}{\tbase-1}
        \cdot \tbase^{\mlevel{\parent(s)}} > \eps \cdot
        d_\MTR(\rep_{\parent(s)}, \rep_t)$.
        $\frac{2\tbase}{\tbase-1} \cdot \tbase^{\mlevel{s}} \leq
        \eps \cdot d_\MTR(\rep_{s}, \rep_t)$.
    \end{enumerate}
\end{lemma}

\begin{proposition}
    The while loop finds a pair in $W$ after $O(\log \sAprx)$ steps.
\end{proposition}
\begin{proof}
    Denote by $\brc{u_0,v_0}$ the pair with which loop begin with.  It
    is straightforward to see that the loop climb through all ancestor
    pairs $\brc{u,v}$ of $\brc{u_0,v_0}$ that satisfy either (i)
    $\mlevel{u}\leq \mlevel{v}<\mlevel{\parent(u)}$, or (ii)
    $\mlevel{u}< \mlevel{v}=\mlevel{\parent(u)}$ and
    $\parent(u)\preceq v$.
    
    Thus, if exists an ancestor pair in $W$, it will be found by the
    loop.  As we argue in \lemref{WSPD} there exists an ancestor pair
    $\brc{\bar{u},\bar{v}}$ of $\brc{x,y}$ in $W$.  Our choice
    $\brc{u_0,v_0}$ ensures that $u_0$ is descendant of $\bar{u}$ at
    most $O\pth{ \log \sAprx }$ levels down $T$, and the same goes for
    $v_0$ and $\bar{v}$.
\end{proof}

Combining the above claims, implies the following:
\begin{lemma}
    Let $P$ be a $n$-point metric.  Assume we are given a
    $(\sP,\sS,\sQ,\sAprx)$-CRS $\CA$
    of a set ${P}$, where $\sAprx \leq 3n^2$. Then, one can obtain
    $(\sP',\sS',\sQ',1+\eps)$-CRS $\CB$
    of ${P}$, where $\sP'=\sP + 2^{O(\dim)}n \log n +
    \eps^{O(\dim)}n$, $\sS'= \sS+ \eps^{-O(\dim)} n$,
    $\sQ'= \sQ + O(\log \sAprx)$.
    
    \lemlab{crs-boost}
\end{lemma}

\begin{remark} 
    The dependence of the query time on $\sAprx$ can be improved from
    $O(\log \sAprx)$ to $O(\log \log \sAprx)$ without sacrificing any other
    parameter. The idea is to replace the ``ladder climbing'' in the
    algorithm above (the while loop) with a binary search on the $\log
    \sAprx$ levels. To do so we change the \WSPD{} procedure to output
    \emph{all} pairs it encounters. This clearly does not change
    asymptotically the size of $W$. We do a binary search on the $\log
    \sAprx$ relevant levels to find the lowest level pairs which still
    appear in the WSPD, and this gives as the relevant pairs. We do
    not pursue this improvement rigorously, since in the CRS we
    develop in the next section, the query time $\sQ$ dominates
    $\sAprx$ anyway, and thus this would lead to no asymptotic saving
    in the query time.
    
    \remlab{log:log:A}
\end{remark}

\subsection{Assouad Embedding}
\seclab{do:Assouad}

To obtain a constant approximation of the distance quickly, we will use
a theorem due to Assouad~\cite{a-pldr-83} (see also
\cite{h-lams-01,gkl-bgfld-03}). The following is a variant of the
original statement, tailored for our needs, and its proof is provided
for the sake of completeness.

\begin{theorem} 
    Any metric space $\MTR$ with doubling dimension $\dim$, can be
    embedded in $\ell_\infty^d$, where $d\leq \eps^{-O(\dim)}$, such
    that the metric $(\MTR, \sqrt{d_\MTR})$ is distorted by $1+\eps$
    factor.
    
    \theolab{Assouad}
\end{theorem}
% As we shall discuss later, quantitively improved theorem is obtained
% in \cite{gkl-bgfld-03}.  For completeness we now present a proof.

\begin{proof}
    Fix $r>0$, we begin by constructing an embedding $\phi^{(r)}: \MTR
    \rightarrow \Re^{d_1}$, where $d_1=\eps^{-O(\dim)}$ with the
    following properties: For every $x,y\in\MTR$:
    \begin{enumerate}
        \item $\dist{\phi^{(r)}(x) - \phi^{(r)}(y)}_\infty \leq
        \min\brc{ r, d_\MTR(x,y)}$.
        
        \item If $d_\MTR(x,y)\in [\,(1+\eps)r, 2r)$ then $\dist{\phi^{(r)}(x) -
           \phi^{(r)}(y)}_\infty \geq (1-\eps) r$.
    \end{enumerate}
    
    We take an $\eps r$-net $\CN^{(r)}$ of $\MTR$ and color it such
    that every pair $x,y\in \CN^{(r)}$ for which $d_\MTR(x,y) \leq
    4r$, is colored differently.  Clearly, $d_1=\eps^{-O(\dim)}$
    colors suffices. Associate with every color $i$ a coordinate, and
    define for $x\in \MTR$, $\phi^{(r)}_i(x)= \max\sbrc{0, r -
       d_\MTR(x,C_i)}$, where $C_i\subset \CN^{(r)}$ is the set of
    points of color $i$.
    
    We next check that the two properties above are satisfied. As
    $\phi^{(r)}_i(x)\in[0,r]$, it is clear that $|\phi^{(r)}_i(x) -
    \phi^{(r)}_i(y)| \leq r$, for every color $i$.  The 1-Lipschitz
    property easily follows from the triangle inequality.
    
    Next, assume that $d_\MTR(x,y)\in[\,(1+\eps)r,2r]$. Since
    $d_\MTR(x,\CN^{(r)}) \leq \eps r$, there exists a color $i$ for
    which $d_\MTR(x,C_i) \leq \eps r$. This implies (by the
    triangle inequality) that $d_\MTR(y,C_i)\geq r$, hence
    $|\phi^{(r)}_i(x)-\phi^{(r)}_i(y)| \geq (1-\eps) r$.  Hence, the
    concatenation of all these coordinates, $\phi^{(r)}=\oplus_i
    \phi^{(r)}_i$ satisfies the condition above.
    
    Let $d_2=8\eps^{-1}\log(\eps^{-1})$.  The final embedding
    $\phi:\MTR \rightarrow \Re^{d_2 d_1}$ is done by combining a
    weighted sum of $\phi^{(r)}$ as follows.  Let $M_l(x)$ denote the
    matrix of size $d_2\times d_1$, such that it is all zero, except the
    $(l \pmod{d_2})$th row, which is $\psi_{l}(x) =
    \phi^{((1+\eps)^l)}(x)$.  Then
    \[ 
    \phi(x)= \sum_{l\in\ZZ} \frac{M_l(x)}{(1+\eps)^{l/2}}.
    \]
    
    To see that the embedding is $1+O(\eps)$ approximation of
    $\sqrt{d_\MTR}$, fix a pair of points $x,y\in \MTR$, and let $l_0
    \in \ZZ$ such that $d_\MTR(x,y)\in [(1+\eps)^{l_0+1}, 
    (1+\eps)^{l_0+2})$.  Then in the relevant coordinates the $\ell_\infty$
    distance between $x$ and $y$ is 
    \begin{multline*}
        \dist{\sum_{k \in \ZZ} \pth{ \psi_{l_0+d_2 k }(x) - \psi_{l_0+
                 d_2k }(y)}
        }_\infty 
        \geq \dist{ \psi_{l_0}(x) - \psi_{l_0}(y) }_\infty - \sum_{k
           <0} \dist{\psi_{l_0+d_2 k}(x) - \psi_{l_0+d_2 k}(y)
        }_\infty \\
        - \sum_{k>0} \dist{ \psi_{l_0+d_2 k}(x)
           - \psi_{l_0+d_2 k}(y) }_\infty \\
        \geq (1-\eps) (1+\eps)^{l_0/2} - \sum_{k<0} \frac{
           (1+\eps)^{2+l_0+d_2 k}}{(1+\eps)^{(l_0+d_2k)/2}} -
        \sum_{k>0} \frac{
           (1+\eps)^{2+l_0}}{(1+\eps)^{(l_0+d_2 k)/2}} \\
        \geq (1-\eps) \cdot (1+\eps)^{l_0/2} - \eps \cdot
        (1+\eps)^{l_0/2} - \eps \cdot (1+\eps)^{l_0/2} \geq
        (1-O(\eps)) \sqrt{d_\MTR(x,y)}.
    \end{multline*}
    
    On the other hand, for each $j\in \brc{0,\ldots,d_2-1}$,
    \begin{multline*}
        \dist{\sum_{k \in \ZZ} (\psi_{l_0+j+d_2 k}(x) -
           \psi_{l_0+j+d_2k}(y))
        }_\infty \\
        \leq \sum_{k \leq0} \dist{\psi_{l_0+j+d_2 k}(x) -
           \psi_{l_0+j+d_2 k}(y) }_\infty + \sum_{k>0} \dist{
           \psi_{l_0+j+d_2k}(x)
           - \psi_{l_0+j+d_2 k}(y) }_\infty \\
        \leq \sum_{k\leq 0}
        \frac{(1+\eps)^{2+l_0+j+d_2k}}{(1+\eps)^{(l_0+j +d_2 k)/2}} +
        \sum_{k>0} \frac{ (1+\eps)^{2+l_0}}{(1+\eps)^{(l_0+j+d_2
              k)/2}} = (1+O(\eps)) \sqrt{d_\MTR(x,y)}.
    \end{multline*}    
    Hence $\dist{\phi(x)-\phi(y)}_\infty$ is $1+O(\eps)$ approximation
    to $\sqrt{d_\MTR(x,y)}$.
\end{proof}

The relevance of Assouad's embedding to compact representations is
clear: Intuitively, $\phi(x)$ is short, and given $\phi(x)$ and
$\phi(y)$, we can compute the square of the $\ell_\infty$ norm of the
difference, and obtain $1+\eps$ approximation to $d_\MTR(x,y)$. Note
however, that in order to be able to do it, we need to store
$\Theta(\log (\Spread/\eps))$ bits for each real number, which may
require many words to be represented in our computation model
(see~\secref{preliminaries}). We solve this issue in \lemref{red-poly}
by reducing the problem for metrics with arbitrary spread a to a set
of similar problems on metrics with only polynomial spread, on which
Assouad's embedding can be applied.

\begin{lemma} 
    Given $n$-point metric $M$ with a polynomially bounded spread
    $\Spread$ and doubling dimension $\dim$, an Assouad's embedding 
    (with parameter $\eps$) of
    $M$ can be computed in $\eps^{-O(\dim)}n \log^2n$ time.
    
    \lemlab{Assouad:alg}
\end{lemma}
\begin{proof}
    We follow closely the proof of \theoref{Assouad}. For each scale
    $(1+\eps)^l$, we find a $\eps(1+\eps)^l$-net $\CN^{((1+\eps)^l)}$
    from the \nettree{} which is $O(\eps (1+\eps)^l)$ cover and
    $\Omega(\eps (1+\eps)^l)$ separated in $O(n)$ time.  We define a
    graph on this net: two points are connected by an edge if they are
    at distance at most $4 (1+\eps)^l$. 
    This can be done in $\eps^{-O(\dim)}n$ time using a
    variant of $\Rel{}$ sets (basically, we compute sets like
    $\Rel{}$ that contain points at distance at most $O(\eps^{-1})$ times
    the current scale, instead of
    $13$ times the current scale).
    We then partition $\CN^{((1+\eps)^l)}$ to color-classes using the greedy
    algorithm. Implemented with hashing, it works in expected $O(n)$
    steps.  Next, for each color class we construct a $(1+\eps/2)$-ANN
    data structure, and we thus can compute an
    $(1+\eps/2)$-approximation to $d_\MTR(x,C_i)$. Note that in the
    proof of \theoref{Assouad}, by enlarging the constants a little
    bit, $(1+\eps/2)$-approximation suffices.  We repeat this
    construction for the $\log_{1+\eps} \Spread$ levels in the metric.
    The rest of the embedding calculation is straightforward.
    
    The running time of the algorithm is therefore $\eps^{-O(\dim)} n \log
    n \log \Spread$.
\end{proof}

\begin{remark}
    We believe that for $\eps=100$, a similar embedding can be
    constructed directly on the \nettree{} in $2^{O(\dim)} n$ time.
    The construction seems however much more complicated than the one
    described in \lemref{Assouad:alg}. We have therefore decided that
    the slight gain in preprocessing time (overall, a factor of $\log
    n$, since the running time for constructing the \nettree{} is
    $2^{O(\dim)}n \log n$) does not worth the complications.
\end{remark}

\begin{lemma} 
    If there exists a $(\sP,\sS,\sQ,\sAprx)$-CRS $\CA$ for an
    $n$-point metric with doubling dimension $\dim$ and spread $\leq 3
    (n/\eps)^{12}$, and if $\sP$ is concave. Then there exists
    $(P(4n)+2^{O(\dim)} n \log n, S+O(n),Q+O(1),(1+\eps)\sAprx)$-CRS $\CB$
    for finite $\dim$-doubling dimensional metrics, without any
    assumption on the spread.
    
    \lemlab{red-poly}
\end{lemma}
\begin{proof}
    Denote by $H$ the low quality HST of \secref{HST} which is $3n^2$
    approximation to the given metric $\MTR$.
    
    Set $a_1=0$ and $a_2=\ceil{5(\log(\eps^{-1})+ \log_2 n)}$. Apply
    the following procedure on $H$ to obtain new HSTs $H_i$, $i\in
    \brc{1,2}$.  Scan $H$ top down. Retain the root, the leaves, and
    all internal vertices $u\in H$ with the following property: there
    exists $b>0$ such that $\log_2 b\equiv a_i
    \pmod{\ceil{10(\log(\eps^{-1})+ \log_2 n)}}$ and
    $\Delta_{\parent(u)}> b \geq \Delta_u$. The HST $H_i$ is
    constructed naturally on the retained vertices: A retained vertex
    $u$ is connected to a parent $v$ in $H_i$ if $v$ is the lowest
    retained ancestor of $u$ in $H$.
    
    Next, for each non-leaf vertex $u\in H_i$, $i\in\brc{1,2}$, denote
    by $C(u)$ the set of children of $u$. We observe that
    $R(C(u))=\brc{\rep_u |\; u\in C(u)}$ has a spread at most $3
    (n/\eps)^{12}$.  To see this, note that $\diam(R(C(u)))\leq
    \Delta_u$, and on the other hand let $b$ the largest real number
    such that $b<\Delta_u$, and $\log b \equiv a_i
    \pmod{\ceil{10(\log(\eps^{-1}) \log_2 n)}}$.  Obviously $b \geq
    \Delta_u/ (n/\eps)^{10}$ and for every $x,y\in C(u)$,
    $\Delta_{\lca{}_H(x,y)}\geq b$, and therefore $d_\MTR(x,y)\geq
    b/(3n^2)$.  Thus for each internal vertex $u\in H_i$ we can
    construct a ${\sAprx}$-approximate CRS $\CA$ to $R(C(u))$. The
    whole construction time is therefore $2^{O(\dim)}n \log n +\sum_k
    P(n_k) \leq 2^{O(\dim)}n \log n +P(4n)$.
    
    We equip $H$, $H_1$ and $H_2$ with a data structure for handling
    queries for least common ancestor and finding an ancestor at a
    given depth, both in constant time.
    
    A distance query for the pair $x,y\in \MTR$ is processed as
    follows.  Let $u_i=\lca_{H_i}(x,y)$.  let $x_i$ be a child of
    $u_i$ which is an ancestor of $x$ in $H_i$, and similarly $y_i$.
    Note that $u_i,x_i,y_i$ can be computed in constant time using the
    $\lca$ and depth ancestor queries.
    
    Further observe that $\exists i\in\brc{1,2}$ for which
    $\max\sbrc{\Delta_{x_i},\Delta_{y_i}} \leq
    \Delta_{\lca_H(x,y)}/(n/\eps)^5$, and finding this $i$ is an easy
    task.
    
    We next query the CRS $\CA$ of $R(C(u_i))$ for
    approximation of $d_\MTR(\rep_{x_i},\rep_{y_i})$.  From the above
    we deduce that
    \[ 
    \max\sbrc{d_\MTR(x,\rep_{x_i}),d_\MTR(y,\rep_{y_i})} \leq
    \frac{3\eps^5}{n^3} \cdot \frac{\Delta_{\lca_H(x,y)}}{3n^2}\leq
    \frac{3\eps^5}{n^3} d_\MTR (x,y).
    \] 
    and therefore we have obtained $\sAprx (1+\eps)$ approximation to
    $d_\MTR(x,y)$.
\end{proof}

\begin{corollary} 
    Every $n$ point metric with doubling dimension $\dim$ has
    $(\sP=\eps^{-O(\dim)}n \log^2$ $n$, $\sS=\eps^{-O(\dim)} n$,
    $\sQ=\eps^{-O(\dim)}, \sAprx=1+\eps)$-CRS.

    \corlab{coarse-crs}
\end{corollary}
\begin{proof}
    Combine of \lemref{red-poly} and \lemref{Assouad:alg}.
\end{proof}

Note that in \corref{coarse-crs} the query time depends on $\eps$,
in contrast to the claim in \theoref{CRS}~(a). This can be remedied using
\lemref{crs-boost}:

\begin{proof}[Proof of \theoref{CRS} (\ref{item:crs_a})]
    Use the CRS of \corref{coarse-crs} \emph{with constant
       $\eps_0=0.1$} as the bootstrapping CRS in \lemref{crs-boost}.
 \end{proof}
    
\begin{proof}[Proof of \theoref{CRS} (\ref{item:crs_b})]
    In \cite{gkl-bgfld-03}, an alternative proof for Assouad Theorem
    is given with much improved bound on the dimension of the host
    space: They prove that for any metric $(\MTR,d_\MTR)$ with
    doubling dimension $\dim$, it is possible to embed
    $(\MTR,d_\MTR^{1/2})$ in $\ell_\infty^{O(\dim )}$ with distortion
    $O(\dim^2)$.%
    \footnote{If one wants to optimize the distortion using their
       technique, then it is possible to obtain $O(\dim)$ distortion
       when embedding into $\ell_p^{O(\dim \log \dim)}$. }
    
    This embedding can be done in polynomial time. Using it as a
    replacement for \lemref{Assouad:alg}, we therefore obtain the
    claimed CRS.
\end{proof}

\begin{remark}
    The distortion of embedding into $\text{poly}(\dim)$
    dimensional normed space can not be improved below $1.9$, since
    such an embedding gives $1.9$ approximate CRS which use only $O(n
    \poly(\dim) \log \phi)$ bits of storage with label length which
    are polynomially dependent on $\dim$ (see \secref{dist-label}),
    but Talwar~\cite{t-beald-04} have shown that such CRS necessarily
    use at least $n 2^{\Omega(\dim)}$ bits, which is impossible for
    $\dim=\Omega(\log \log n)$.  In this sense the embedding technique
    of \cite{gkl-bgfld-03} can not replace Assouad's original
    technique.
\end{remark}

It is still open whether the construction time in \theoref{CRS}~(b)
can be improved to near-linear. The difficulty lies in the algorithmic
version of the Lov\'asz Local Lemma. As discussed in \remref{log:log:A},
distortions as high as $2^{2^{O(\dim)}}$ are tolerable in this
context.

\subsubsection{Lower Bound}

We next argue that beating the $\Omega(\dim)$ query time using schema
similar to the one presented above, is unlikely.

For given reals $d_1,D,d_2>1$, we say that
$(d_1,D,d_2)$-\emph{Assouad-type-scheme} (ATS) exists if there is a
monotone increasing bijection $f:[0,\infty)\to [0,\infty)$, such that
for all finite metric spaces $(P,d_\MTR)$, with doubling dimension at
most $d_1$, there exists $\phi:P\to
\mathbb{R}^{d_2}_{\dist{\cdot}_X}$, such that for $ x,y\in P$, we have
\[ 
\frac{d_\MTR(x,y)}{D} \leq f\pth{\dist{\phi(x)-\phi(y)}_X } \leq
d_\MTR(x,y).
\]

For example, the embedding of \cite{gkl-bgfld-03} cited above is
$(d_1,O(d_1^2),O(d_1))$-ATS for any $d_1>1$, and it uses $f(x)=x^2$.

\begin{proposition}
    If $d_2\leq d_1/5$, then for any $D>1$, no $(d_1,D,d_2)$-ATS
    exists.
\end{proposition}
\begin{proof}
    The argument distinguishes between two essential cases:
    ``Concave'' function $f$ can not be used in any ATS since it
    causes a violation of the triangle inequality.  For ``convex''
    functions $f$ we slightly generalize an argument from
    \cite{bdgrrr-aaeild-05} that uses topological considerations
    (Borsuk-Ulam theorem) to conclude the impossibility.
    
    Indeed, fix a $(d_1,D,d_2)$-ATS with a function $f$, where
    $d_2\leq d_1/5$.  Denote by $g:[0,\infty)\to [0,\infty)$ where
    $g=f^{-1}$.
    
    Suppose first $\sup_{0<a\leq b<\infty} \frac{g(b)/b }{g(a)/a}
    =\infty$ (``concave $f$'').  Fix $a$ and $b$, such that $0<a<
    b<\infty$ and $\frac{g(b) /b }{g(a)/a } \geq 100 D$.  Let
    $n=\ceil{2Db/a}$, and let $P$ be the line metric on
    $\{0,\ldots,n\}$ such that $d_\MTR(i,j)=a|i-j|$.  By the
    assumption, there exists $\phi:P\to
    \mathbb{R}^{d_1}_{\dist{\cdot}_X}$, such that
    $\dist{\phi(i)-\phi(i+1)}_X \leq g(d_\MTR(i,i+1))=g(a)$, and on
    the other hand
    \[
    g(b) \leq g\pth{ \frac{\ceil{2Db/a} a}{D} } = g\pth{
       \frac{d_\MTR(0,n)}{D}}
    \leq
    \dist{\phi(0)-\phi(n)}_X,
    \]
    since $g$ is monotone increasing, as $f$ is monotone increasing.
    Then by the triangle inequality
    \[
    g(b)\leq \dist{\phi(0)-\phi(n)}_X \leq \sum_{i=1}^n
    \dist{\phi(i-1)-\phi(i)}_X \leq n\, g(a) \leq 4D\, \frac{b\,
       g(a)}{a},
    \]
    which implies that $\frac{g(b)/b}{g(a)/a} \leq 4D$, which is a
    contradiction.
    
    Next, assume that there exists $C>1$ such that $\sup_{0<a\leq
       b<\infty} \frac{g(b) \cdot a}{g(a)\cdot b} \leq C$ (``convex
    $f$''). Then, for any $a \leq b$ we have $\tfrac{g(b) a}{C b} \leq
    g(a)$.  In particular, we have $\tfrac{g(d_\MTR(x,y))
       (d_\MTR(x,y)/D)}{C d_\MTR(x,y)} \leq g(d_\MTR(x,y)/D)$. Namely,
    \[ 
    \frac{g(d_\MTR(x,y))}{C\cdot D} \leq g\pth{\frac{d_\MTR(x,y)}{D}}
    \leq \dist{\phi(x)-\phi(y)}_X \leq g(d_\MTR(x,y)) .
    \] 
    Since $\dist{\cdot}_X$ is $d_2$ dimensional, by John's theorem
    (see \cite[Ch.~V]{b-acc-02}) it can be approximated by
    $\dist{\cdot}_2$ up to a $\sqrt{d_2}$ factor.  We thus have a
    $C'>1$ such that for any $d_1$ dimensional finite metric
    $(P,d_\MTR)$, there exists $\phi':(P,d_\MTR)\to \Re^{d_2}_{\dist{\cdot}_2}$
    satisfying
    \begin{equation} \label{eq:gd-phi'}
        \frac{g(d_\MTR(x,y))}{C'} 
        \leq \dist{\phi'(x)-\phi'(y)}_2 \leq g(d_\MTR(x,y)) .
    \end{equation}
    
    We next estimate how much $g\circ d_\MTR$ distorts $d_\MTR$ as a
    function of the spread of $P$. Assume that $\min_{x\neq y \in
       P}d_\MTR(x,y)=a_1$, and $\max_{x\neq y \in P}d_\MTR(x,y)=b_1$,
    that is $\Spread(P) = b_1/a_1$.  Then
    \begin{align*}
        \max_{a_1\leq t}\frac{g(t)}{t} &=\frac{g(a_1)}{a_1} \cdot
        \max_{a_1\leq t} \frac{g(t) a_1}{t g(a_1)} \leq C \frac{g(a_1)}{a_1},\\
        \text{and ~~~}\max_{s\leq b_1}\frac{s}{g(s)} &=\frac{b_1}{g(b_1)} \cdot
        \max_{s\leq b_1} \frac{g(b_1) s}{b_1 g(s)} \leq C
        \frac{b_1}{g(b_1)}.
    \end{align*}
    Thus, consider the ``distortion''  of $g$, we have
    \[
    \frac{\max _{x\neq y \in P}\frac{g(d_\MTR(x,y))}{d_\MTR(x,y)}
       \cdot \max _{x\neq y \in P}\frac{d_\MTR(x,y)}{g(d_\MTR(x,y))}
    }{\Spread(P)} \leq \frac{ C \frac{g(a_1)}{a_1} \cdot
       C \frac{b_1}{ g(b_1)} }{b_1/a_1} =
    C^2 \,\frac{g(a_1)}{g(b_1)} .
    \]
    As $g(0)=0$ and $\lim_{x\to \infty}g(x)=\infty$ we conclude that
    this ratio tends to $0$ as the spread $\Spread(P)$ tends to
    $\infty$.  Combining it with \eqref{eq:gd-phi'}, we conclude that
    for $\hphi:(P,d_\MTR) \to \Re^{d_2}_{\dist{\cdot}_2}$, defined as
    $\hphi(x)=\phi'(x)$,
       have
    $\text{dist}(\hphi)=o(\Spread(P))$.  We will next show that this
    is impossible when $P$ is sufficiently dense net of
    $\mathbb{S}^{d_2}$.
    
    Let $0<\eta\leq 0.1$.  We take $P=P_\eta$ to be a $\eta$-net of
    $\mathbb{S}^{d_2}_{\dist{\cdot}_2}$.  The finite metric $P_\eta$
    has doubling dimension at most $d_1$.  From the above we can embed
    $\phi':P_\eta\to \Re^{d_2}_{\dist{\cdot}_2}$ with distortion
    $o(\Spread(P)) = o(\eta^{-1})$. By scaling we may assume that this
    embedding is 1-Lipschitz.  By Kirszbraun Theorem (see
    \cite[Ch.~1]{bl-gnfa-00}), the embedding $\phi'$ can be extended
    to the whole sphere $\hphi':\mathbb{S}^{d_2}_{\dist{\cdot}_2}\to
    \mathbb{R}^{d_2}_{\dist{\cdot}_2}$ without increasing the
    Lipschitz constant.  Borsuk-Ulam theorem (cf.  \cite{m-ubut-03})
    states that there exists $x\in\mathbb{S}^k$ such that
    $\hphi'(x)=\hphi'(-x)$.  Note that $\exists y,z\in P_\eta$ such
    that $\dist{x-y}_2\leq \eta$, and $\dist{(-x)-z}_2\leq \eta$.
    Since $\hphi'$ is $1$-Lipschitz, we have
    \[ 
    \dist{\phi'(y)-\phi'(z)}_2=
    \dist{\hphi'(y)-\hphi'(z)}_2 \leq
    \dist{\hphi'(y)-\hphi'(x)}_2+
    \dist{\hphi'(-x)-\hphi'(z)}_2 \leq 2 \eta .
    \] 
    On the other hand $\dist{y-z}_2\geq 1-2\eta$, which means that the
    Lipschitz constant of $\phi'{}^{-1}$, and thus the distortion
    of $\phi'$, is at least $\Omega(\eta^{-1})$.  This is a
    contradiction for sufficiently small $\eta>0$, since we argued
    above that the distortion must be $o(\Spread(P)) = o(\eta^{-1})$.
\end{proof}

\subsection{Distance Labeling}
\seclab{dist-label}

Approximate distance labeling scheme (ADLS) seeks to compute for each
point in the metric a short label such that given the labels of a pair
of points, it is possible to compute efficiently an approximation of
the the pairwise distance. Thus, ADLS is a stricter notion of compact
representation.\footnote{When comparing the storage of ADLSs to that
   of the CRSs from the previous sections, note that here we count
   \emph{bits}, whereas in the rest of the paper we count \emph{words}
   of length $O(\log n+ \log\log \Spread +\log \eps^{-1})$.}  This
notion was studied for example in
\cite{gppr-dlg-04,gkkpp-adls-01,tz-ado-01}.

In the constant doubling dimension setting Gupta \etal{]
\cite{gkl-bgfld-03} have shown an $(1+\eps)$-embedding of the metric
in $\ell_\infty^{O(\log n)}$.  This implies $(1+\eps)$-ADLS with
$O(\log n \log \Spread)$ bits for each label (the $O$ notation here
hides constants that depend on $\eps$ and $\dim$). Talwar
\cite{t-beald-04} has shown an improved $(1+\eps)$-ADLS with only
${\eps}^{-O(\dim)} \log {\Spread}$ bits per label.  Slivkins
\cite{s-date-05} has shown a $(1+\eps)$-ADLS with $\eps^{-O(\dim)}
\log^2 n \log\log \Spread $ bits per label.  Their techniques seem to
be very different from each other.

Here we improve Slivkins' result and unify it with Talwar result under
the same technique.

\begin{proposition}
    Given a finite metric space, one can build a $(1+\eps)$-ADLS with
    \[\min\sbrc{ \eps^{-O(\dim)} \log \Spread, \eps^{-O(\dim)} \log n
       (\log n+ \log\log \Spread )}\] bits per label.
    
    Furthermore, there exist one dimensional finite metric spaces of
    size $n$, and spread $\Phi \geq 2^{2n}$ for
    which any $1.9$-ADLS requires labels of size $\Omega(
    \log n \log \log \Spread)$ bits per label.
\end{proposition}
\begin{proof}[sketch]
    First, labels of length $\eps^{-O(\dim)} \log \Spread$ follow
    directly from \theoref{Assouad}: We have $\eps^{-O(\dim)}$
    coordinates, and, as discussed after the proof of
    \theoref{Assouad}, we only need $O(\log(\Spread /\eps))$ bits of
    accuracy for each coordinate.

    We next show $(1+\eps)$ approximate distance labeling scheme using
    $\eps^{-O(\dim)}\log n ( \log n +\log\log \Spread )$ bits per
    label.  We do so by presenting a ``distributed implementation'' of
    the data-structure used to prove \corref{coarse-crs}. That data
    structure consists of two trees (HSTs) $H_1,H_2$ on the same set
    of leaves: the points of the metric.  Given two points $x^1,x^2$,
    we compute $u_i=\lca_{H_i}(x^1,x^2)$, and $x^j_i$ the ancestor of
    $x^j$ in $H_i$ which is the child of $u_i$.  We then apply an
    Assouad embedding $A(x^j_i)$ that uses $O(\log n + \log \log
    \Spread +\log (\eps^{-1}))$ bits. We define an identifier $I(v)$
    of vertex $v\in H_i$ to be $A(v)$ concatenated with the $\Delta_v$
    (encoded with $O(\log \log \Spread )$ bits). Hence, given two
    points $x^1,x^2$, using the identifiers $I(x^1_1)$, $I(x^2_1)$,
    $I(x^1_2)$, $I(x^2_2)$, $I(u_1)$, $I(u_2)$, we can compute
    $1+\eps$ approximation of $d_\MTR(x_1,x_2)$. We now use (the proof
    of) a result of Peleg~\cite{p-ilsg-04}: Given an $n$-vertex rooted
    tree with identifiers $I(v)$ of maximum length $s$ on the
    vertices, it is possible to efficiently compute labels $L(v)$ of
    length $O(\log n (\log n +s))$ to the vertices, such that given
    $L(x)$ and $L(y)$ one can efficiently decode $I(u)$, where
    $u=\lca(x,y)$.

    Unfortunately, we need a little bit more: an access to the
    children of $u$ which are the ancestors of $x$ and $y$.  In order
    to achieve it we tinker with the construction of Peleg: In
    Definition 3.2 in \cite{p-ilsg-04}, we extend the tuple $Q_i(v)$
    to be
    \begin{equation*}
        Q_i(v)= \biggl \langle 
        \langle i-1,I(\gamma_{i-1}(v)) \rangle,
        \langle i,I(\gamma_{i}(v)) \rangle,
        \langle i+1,I(\gamma_{i+1}(v)) \rangle,
        \underline{\langle i,I(\mathrm{hs}(\gamma_{i}(v))) \rangle}
        \biggr \rangle,
    \end{equation*}
    where $\mathrm{hs}(u)$ is the \emph{heavy sibling} of $u$ (the
    underlined part is our extension).  By studying Peleg's
    construction, it is easy to verify that this extension suffices.

    The above construction is asymptotically optimal in terms of $n$
    and $\Phi$ when $\Phi\geq 2^{2n}$, as we now prove.  In
    \cite{gppr-dlg-04} a family of $n$-vertex weighted rooted binary
    trees, such that any exact distance labeling scheme of the
    \emph{leaves} requires labels of length $\Omega(\log n \log M)$
    bits, where the edge weight is in the range $\{0,\ldots,M-1\}$.  A
    further property of that family of trees is that the depth
    $h=M\log_2 n$ (i.e., the distance from the root) of all the leaves
    is the same.  We next transform each tree $T$ in that family into
    an HST $H$ by giving every vertex $v$ a label
    $2^{-\text{depth}_T(v)}$. For any two leaves $x$ and $y$ let
    $d_T(x,y)= 2(h+\log_2 d_H(x,y))$.  Furthermore, even $1.9$
    approximation of $d_H(x,y)$ allows us to recover the exact value
    of $d_H(x,y)$, since this value is an integral power of $2$.  Let
    us summarize: Given a $1.9$ approximation of the distance in $H$
    allows us to obtain the exact distance in $T$. Therefore by
    setting $M=(\log_{2} \Phi) / n$, it proves a lower bound of
    $\Omega( \log n \log \log \Spread )$ on the average label's length
    for $1.9$-ADLS for this family of HSTs.  Since these HSTs are
    binary their doubling dimension is $1$.
\end{proof}

After a preliminary version of this paper appeared, Slivkins
\cite{s-deolr-05} managed to produce an ADLS with labels length of
$\eps^{-O(\dim)} \log n \log \log \Phi$, which improves upon our
construction in the range $n^{\log \log n} \ll \Phi \ll 2^n$.

%-----------------------------------------------------------------
%-----------------------------------------------------------------
%-----------------------------------------------------------------

\section{Doubling Measure}
\seclab{doubling:measure}

A measure $\mu$ on a metric space $\MTR$ is called $\eta$-doubling 
if for any $x\in \MTR$ and $r\geq 0$,
$\mu(\Ball(x,2r))\leq \eta \cdot \mu(\Ball(x,r))$.  Doubling measure
is already a useful notion in analysis of metric spaces (see
\cite{h-lams-01}), and has recently been used in some algorithmic
applications \cite{s-date-05}.  Vol$'$berg and Konyagin
\cite{vk-omdc-87} proved that any compact $\lambda$ doubling metric
space has $\lambda^{O(1)}$-doubling measure (the opposite direction is
easy).  Wu's proof of this theorem \cite{w-hddmms-98} can be implemented
in linear time on the \nettree{} (for finite metric spaces).  

We assume that the \nettree{} $T$ is already given. Denote by $\deg(v)$
the number of children of $v\in T$. Let $\gamma=\max_{v\in
   T}\deg(v)$ be the maximum degree in $T$. As we have seen before,
$\gamma\leq 2^{O(\dim)}$.  The probability measure $\mu$ is
computed by calling to \Partition{}$(\Root,1)$, where
\textsf{Partition} is defined recursively as follows. 

\begin{center}
    \fbox{
       \begin{minipage}{6cm}
           \begin{tabbing}
               \ \ \ \ \= \ \ \ \ \ \= \ \ \ \ \= \kill
               \Partition{}$(u\in T, p_u\in [0,1])$. \+ \\
               \If $u$ is a leaf \Then \+\\
               Set $\mu(\brc{\rep_u}) \leftarrow p_u$. \\
               \< \Else\\
               \For each child $v$ of $u$ with $\rep_v\neq \rep_u$
               \Do\\
               \> Set $p_v\leftarrow p_u / \gamma$.\\
               \> Call \textsf{Partition}$(v,p_v)$. \\
               Let $v_0$ be the unique child of $u$ 
               such that $\rep_{v_0}=\rep_u$. \\
               Set $p_{v_0} \leftarrow 
               p_u \pth{1 - (\deg(u)-1)/\gamma }$.\\
               Call \textsf{Partition}$(v_0,p_{v_0})$.\- 
           \end{tabbing}
       \end{minipage}}
\end{center}

\begin{claim}
    For any $u\in T$, we have $p_u=\mu(P_u)$.
\end{claim}
\begin{proof}
    By straightforward induction on the height of $T$.
\end{proof}

\begin{claim}
    Fix $l \in \mathbb{N}$, and two vertices $u$ and $v$ in $T$, such
    that $\max\sbrc{\mlevel{u},\mlevel{v}}< l \leq
    \min\brc{\mlevel{\parent(u)}, \mlevel{\parent(v)}}$ and
    $d_\MTR(\rep_u,\rep_v) \leq 40 \tbase^{l}$. Then $p_u \leq
    \gamma^{O(1)} p_v$.
    
    \clmlab{balanced:distrib}
\end{claim}
\begin{proof}
    Denote by $w=\lca_T(u,v)$, and by $w=u_0,u_1,\ldots, u_a=u$
    the path in $T$ from $w$ to $u$, and by $w=v_0, v_1,\ldots,
    v_b=v$ the path in $T$ from $w$ to $v$.
    
    We claim that for any $i\geq 1$, if $\mlevel{u_i}> l+3$, then
    $\rep_{u_i}\neq \rep_{u_{i+1}}$. Indeed, otherwise
    \begin{multline*}
        d_\MTR(\rep_{u_{i}},\rep_v)
        \leq d_\MTR(\rep_{u_{i+1}},\rep_u)+ d_\MTR(\rep_u,\rep_v) \\
        \leq \tfrac{2\tbase}{\tbase-1}\cdot
        \tbase^{\mlevel{u_{i+1}}} + 40\tbase^l \leq
        \tfrac{2}{\tbase-1}\tbase^{\mlevel{u_i}}+ 40\tbase^{-4}
        \cdot \tbase^{\mlevel{u_i}} \leq 
        \tbase^{\mlevel{u_i}}/4,
    \end{multline*}
    but this is a contradiction to the packing property of
    \defref{net:tree}, since $v\notin P_{u_i}$ (note that for this
    argument to work, $\tbase$ need to be large enough constant, say
    11).
    
    Next, we claim that for any $i\geq 1$ for which
    $\mlevel{u_i}> l+3$, $\mlevel{u_{i-1}} =\mlevel{u_i}+1$.
    Otherwise, $\mlevel{u_{i-1}} -1 \geq \mlevel{u_i} + 1$ 
    implying
    \begin{eqnarray*}
        d_\MTR(\rep_{u_{i}},\rep_v) &\leq& 
        d_\MTR(\rep_{u_{i+1}},\rep_u)+ d_\MTR(\rep_u,\rep_v) \leq
         \tfrac{2\tbase}{\tbase-1}
        \cdot \tbase^{\mlevel{u_i}} + 40 \tbase^{-4} \cdot
        \tbase^{\mlevel{u_i}} \\
        &=& (\tfrac{2\tbase}{\tbase
           (\tbase-1)}+ \tfrac{40}{\tbase^5}) \cdot
        \tbase^{\mlevel{u_i}+1} 
        \leq 
        \tbase^{\mlevel{u_{i-1}}-1} / 4 =
        \tbase^{\mlevel{\parent(u_i)}-1} / 4,
    \end{eqnarray*}
    contradicting the packing property of \defref{net:tree}, since
    $v\notin P_{u_i}$.
    
    Thus, the path between $u$ and $w$ is full, containing vertices on
    all levels, except maybe the last three levels. 
    Furthermore, the representatives are different in each level. 
    We therefore conclude that $p_u \leq p_w /
    \gamma^{\mlevel{w}-l-4}$.  On the other hand, $p_v \geq p_w
    \gamma^{\mlevel{w}-l+1}$. Therefore $p_u\leq \gamma^5 p_v$.
\end{proof}

\begin{theorem}
    For any $n$-point metric space having doubling dimension $\dim$ it
    is possible to construct a $2^{O(\dim)}$ doubling measure in
    $2^{O(\dim)} n \log n$ time.
\end{theorem}
\begin{proof}     
    The running time of \Partition{} is clearly linear, and is dominated
    by the time to construct the \nettree{}.

    We are left to prove that $\mu$ is a $\lambda^{O(1)}$-doubling
    measure.  Let $x\in P$ and $r>0$.  Denote by $N=\brc{u\in T \sep{
          \mlevel{u}\leq \log_\tbase (r/8) < \mlevel{\parent(u)}}}$.
    As we have seen in \propref{net:tree}, the representatives of the
    vertices of $N$ forms a net in the right scale. In particular,
    there exists $\hx\in N$ such that $d_\MTR(x,\rep_{\hx}) \leq 3r/8$
    and $P_{\hx}\subset \Ball(\rep_{\hx},3r/8)\subset \Ball(x,r)$.
    Hence $p_{\hx}\leq \mu(\Ball(x,r))$.  On the other hand, any two
    different representatives of vertices from $N$ are at least $r/40$
    separated, and therefore, for $X=N\cap \brc{u\in T \sep{ \rep_u\in
          \Ball(x,3r)}}$, we have $|X| \leq \lambda^{O(1)}$.  Note
    that $\Ball(x,2r)\subset \cup_{u\in X} P_u$, and therefore
     \[ 
     \mu(\Ball(x,2r)) \leq \sum_{u\in X} p_u \leq |X| \max_{u\in X}
     p_u. 
     \]
     By \clmref{balanced:distrib}, $\max_{u\in X} p_u \leq
     \lambda^{O(1)} p_{\hx}$.  We conclude that $\mu(\Ball(x,2r)) \leq
     \lambda^{O(1)} \mu(\Ball(x,r))$.
\end{proof}

We note in passing that algorithm \Partition{} can be programmed in
our computational model since every point gets at least $2^{-O(n \log
   n)}$ measure, which can be easily represented in a floating-point
word of length $O(\log n)$. Moreover, the algorithm has a ``built in''
mechanism to handle rounding error: instead of dividing by $\gamma$,
we can divide by say $2\gamma$ and now rounding errors are
automatically offset in the measure given to $v_0$.

%-----------------------------------------------------------------
%-----------------------------------------------------------------

\section{Lipschitz Constant of Mappings}
\seclab{Lipschitz}

\begin{defn}
    A function $f: (P,\nu) \rightarrow (\MTR,\rho)$ is \emph{$K$-Lipschitz}
    if for any $x, y \in P$, we have $\rho(f(x) ,f(y) ) \leq K \cdot
    \nu(x , y)$.
    
    A point $x \in P$ is \emph{$K$-Lipschitz} if, for any $y \in
    P$, we have $\rho( f(x) ,f(y) ) \leq K \cdot\nu(x , y)$.
\end{defn}

Thus, given a set of points $P \subseteq \Re^d$, and a mapping $f:P
\rightarrow \Re^{d'}$, it is natural to ask how quickly can we compute
the Lipschitz constant for $f$ on the set $P$, and more specifically, to
compute it for every point of $P$.

\subsection{The Low Dimensional Euclidean Case}
\seclab{lipschitz-euclidean}

Here, we consider a mapping $f:P \rightarrow (M,\rho)$, where $P
\subseteq \Re$ of size $n$, and $(M,\rho)$ is an arbitrary metric
space given as a matrix.

\begin{proposition}
    Computing the Lipschitz constant for $f$ on $P$ can be done in
    $O(n \log {n} )$ time.
\end{proposition}
\begin{proof}
    Indeed, let $a,b,c$ be three numbers in $P$, such that $a < b <
    c$. Observe that 
    \begin{eqnarray*}
        \frac{\rho(f(c),f(a))}{c - a }
        &\leq&
        % \frac{\dist{f(c) -f(b) + f(b) - f(a)}}{c - b + b - a }
        % \leq 
        \frac{\rho(f(c),f(b)) + \rho(f(b) , f(a))}{c - b + b - a
        }
        \leq
        \max \pth{ 
           \frac{\rho(f(c) , f(b))}{c - b } \;,\;
           \frac{\rho(f(b) , f(a))}{b - a }
           },           
    \end{eqnarray*}
    since for any $p,q,r,s$ positive numbers such that $p/q \leq r/s$,
    we have $p/q \leq (p+r)/(q+s) \leq r/s$. Thus, the Lipschitz
    constant is realized by a consecutive pair of points in $P$.
    We can therefore sort $P$, and compute the slope for
    every consecutive pair. Clearly, the maximum is the Lipschitz
    constant of $f$.
\end{proof}

\begin{proposition}
    Let $P$ be a set of $n$ numbers on the real line, and let $f:P
    \rightarrow \Re$ be a given mapping. One can compute the Lipschitz
    constant of $f$ on every point of $P$ in $O(n \log^2 n )$ time.
\end{proposition}

\begin{proof}
    Consider the set $Q = \brc{ (p, f(p)) |\;  p \in P}$.  Let $p$
    be a point in $P$, and let $L_p$ be the set of points of $Q$
    strictly to the left of $p$ (according to the $x$-order), and
    $R_p$ the set of points to its right. 
    Denote by $\CH(A)$ the convex hull of $A\subset \Re^2$.
    If we know the tangents to
    $\CH(L_p)$ and $\CH(R_p)$ that passes through $p$, then we can
    compute the Lipschitz constant of $p$ in constant time (i.e., it
    is the slope of the tangent with largest slope).
    
    Here, one can use the data-structure of Overmars and van Leeuwen
    \cite{ol-mcp-81}, which supports the maintenance of convex-hull
    under insertions, deletions and tangent queries in $O( \log^2 n)$
    per operation. Indeed, sort the points of $P$ from left to right.
    Let $p_1,\ldots, p_n$ be the sorted points. 
    Clearly, given $\CH(L_{p_i})$ and
    $\CH(R_{p_i})$ stored in the dynamic convex-hull data-structure, we
    can compute $\CH(L_{p_{i+1}}$ and $\CH(R_{p_{i+1}}$, 
    by deleting $p_{i+1}$ from
    $\CH(R_{p_i})$, and inserting $p_i$ to $\CH(L_{p_i})$. Thus, we can
    compute all the relevant convex-hulls in $O(n \log^2 n)$
    time. Furthermore, when we have $\CH(L_{p_i})$ and $\CH(R_{p_i})$, we
    perform tangent queries to compute the Lipschitz constant of
    $p_i$. Thus, the overall running time is $O(n \log^2 n)$. 
\end{proof}

% \subsection{The two dimensions to one dimension case}

\begin{theorem}
    Given a set $P$ of $n$ points in the plane, and a mapping $f:P
    \rightarrow \Re$, then one can compute the Lipschitz constant of
    $f$ in $O(n \log^2 n)$ expected time.
\end{theorem}

\begin{proof}
Assume that we know that $f$ is $K$-Lipschitz on a set $Q
\subseteq P$, and we would like to verify that it is $K$-Lipschitz on
$\brc{q} \cup Q$, where $q \in P \setminus Q$. This can be visualized
as follows: From every point $p \in P$, there is an associated point
in $\Re^3$, which is $\widehat{p} = (p_x, p_y, f(p))$. Being a
$K$-Lipschitz as far as $p$ is concerned, implies that $q$ must lie
below the upper cone of slope $K$ emanating from $\widehat{p}$, and
above the lower cone of slope $K$ emanating from $\widehat{p}$. Thus,
if we collect all those upper cones, then $q$ must lie below their
lower envelope. However, since the upper cones all have the same slope,
their lower envelope is no more than a (scaled) version of an additive
weighted Voronoi diagram in the plane. Such a diagram can be computed
in $O(n \log {n} )$ time for $n$ points, and a point-location query in
it can be performed in $O(\log n )$ time.

In fact, using the standard Bentley and Saxe technique
\cite{bs-dspsd-80}, one can build a data-structure, where one can
insert such upper cones in $O( \log^2 n)$ amortized time, and 
given a query point $q$ in the plane, decide in $O( \log^2 n)$ which
of the cones inserted lies on the lower envelope vertically above $q$.
Similar data-structure can be build for the upper envelope of the
lower cones.

Thus, if we conjecture that the Lipschitz constant is $K$, then one can
verify it for $P$ in $O( n \log^2 n)$, by inserting the points of $P$
into the upper and lower envelope data-structure described above.
However, let assume that $K$ is too small. Then, after inserting a
subset $Q$ of points into the data-structure, we will try to verify that
the Lipschitz constant for a point $p \in P$ is $K$ and fail. Then, it must be
that the Lipschitz constant of $f$ on $Q \cup \brc{p}$ is realized by
$p$. Thus, we can compute the Lipschitz constant of $p$ in $Q \cup
\brc{p}$ in $O(|Q|)$ time, updated our guess $K$, and rebuild the
upper and lower data-structures for $Q \cup \brc{p}$.

Of course, in the worst case, this would required $O(n^2 \log^2 n)$
running time (i.e., we would fail on every point). However, it is well
known that if we randomly permute the points, and handle the points
according to this ordering, then the value of the Lipschitz constant on
every prefix would change $O(\log n)$ times in expectation. Thus, this would
lead to $O(n \log^3 n)$ expected running time. Moreover, a slightly
more careful analysis shows that the expected running time is $O(n
\log^2 n )$. See \cite{cs-arscg-89} for details of such analysis.
\end{proof}

\subsection{Constant doubling dimension to arbitrary metric}

\begin{theorem}
    Given a metric $(P,\nu)$ of $n$ points having doubling dimension
    $d$, and a mapping $f:P \to (\MTR,\rho)$, where $\MTR$ is an
    arbitrary metric space. Then one can compute $(1
    +\eps)$-approximation of the Lipschitz constant of $f$ in $n
    \eps^{-O(d)} \log^2 n$ expected time.
\end{theorem}

\begin{proof}
    The algorithm:
    \begin{enumerate}
        \item Compute $\eps^{-1}$-\WSPD{} of $P$ according to
        \secref{WSPD}.
        \item Set $K \leftarrow 0$.
        \item For every pair $(A,B)\in \eps^{-1}\text{-WSPD}$ do:
        \begin{enumerate}
            \item Obtain \emph{some} pair of points $a\in A$ and $b\in
            B$.
            \item Compute $K \leftarrow \max \brc{K,
               \frac{\rho(f(a),f(b))}{\nu(a,b)} }$.
        \end{enumerate}
    \end{enumerate}
    
    Obviously the value $K$ computed by the algorithm above is not larger than
    the Lipschitz constant of $f$. We next show that it is not much
    smaller.  Let $x,y\in P$ be a pair in which $f$ obtains its Lipschitz
    constant, i.e., $\frac{\rho(f(x),f(y))}{\nu(x,y)} = \max_{a\neq b}
    \frac{\rho(f(a),f(b))}{\nu(a,b)}$.  Let $\brc{A,B} \in \WSPD{}$
    be a pair such that $x\in A$, $y\in B$.  Our algorithm chooses
    some pair $a\in A$, $b\in B$. Using the triangle inequality we
    have
    \begin{align*}
        % \frac{\rho(f(x),f(y))}{\nu(x,y)} &\ge
        \frac{\rho(f(a),f(b))}{\nu(a,b)} 
        & \ge \frac{\rho(f(x),f(y))-\diam(f(A))-\diam(f(B))}{\nu(x,y)
           +\diam(A)+\diam(B)}\\
        & \ge \frac{\rho(f(x),f(y))-\diam(f(A))-\diam(f(B))}{(1+2\eps)
           \nu(x,y)}
    \end{align*}
    If $\max\sbrc{\diam(f(A)), \diam(f(B))} \leq \eps\cdot
    \rho(f(x),f(y))$ then we conclude that
    \[
    \frac{\rho(f(a),f(b))}{\nu(a,b)}
    \ge \frac{(1-2\eps) \rho(f(x),f(y))}{(1+2\eps) \nu(x,y)} 
    \] 
    and we are done.  Otherwise, assume that $\diam(f(A))> \eps\cdot
    \rho(f(x),f(y))$.  Then there exists $f(a_1),f(a_2)\in f(A)$ for which
    $\rho(f(a_1),f(a_2))> \eps \cdot \rho(f(x),f(y))$, whereas
    \[
    \nu(a_1,a_2)\leq \diam(A) \leq \eps\cdot \nu(A,B) \leq \eps\cdot
    \nu(x,y).
    \]
    So
    \[ 
    \frac{\rho(f(a_1),f(a_2))}{\nu(a_1,a_2)} > \frac{\eps \cdot
       \rho(f(x),f(y))}{\eps\cdot \nu(x,y)},
    \]
    which is a contradiction to the maximality of the pair $\brc{x,y}$.
\end{proof}

%-------------------------------------------------------------------------
%-------------------------------------------------------------------------

\section{Fast approximation of the doubling dimension}
\seclab{doubling-approximation}

\begin{theorem} \theolab{online}
    Given a metric space $\MTR$ with $n$ points, one can approximate
    the doubling dimension $\dim$ of $\MTR$, up to a constant factor, in
    $2^{O(\dim)} n \log n$ expected time.
\end{theorem}

Notice that this theorem, apart from its intrinsic interest, 
also removes the need to specify $\dim$ together with the input
for the other algorithms in this paper.

The algorithm suggested in \theoref{online} naturally uses the \nettree.

\begin{proposition} \proplab{degree}
    Given a \nettree{} $T$ of a metric $\MTR$, and denote by
    $\lambda_T$ the maximum out degree in $T$, then $\log \lambda_T$
    is a constant approximation to $\dim(\MTR)$.
\end{proposition}
\begin{proof}
    Let $v\in T$ be the vertex with the maximum number of children
    $\lambda_T$.  By \defref{net:tree}, any covering of $\Ball(\rep_v,
    \frac{2\tbase}{\tbase -1} \tbase^{\ell(v)})$, by balls of radius
    $\frac{\tbase-5}{4\tbase(\tbase-1)}$ requires at least $\lambda_T$
    such balls. This means that $\dim(\MTR)=\Omega(\log \lambda_T)$.

    The upper bound $\dim(\MTR)=O(\log \lambda_T)$ follows easily from
    the arguments of \secref{doubling:measure}: There, we actually
    prove the existence of $\lambda_T^{O(1)}$-doubling measure in
    $\MTR$, and it easy to prove that the existence of $\alpha$
    doubling measure in $\MTR$ implies that $\dim(\MTR)\leq \alpha$.
\end{proof}

\begin{proof}[Proof of \theoref{online}]
By \propref{degree} it is enough to show an implementation of the algorithm for
constructing the \nettree{} that is oblivious to the the doubling
dimension of the metric. Checking the algorithm in \secref{nets}, we
observe that the algorithms in \secref{Gonzalez:low:spread},
\secref{Gonzalez:high:spread}, and \secref{net:tree} are indeed
oblivious to the doubling dimension. We are therefore left with
describing a doubling dimension oblivious algorithm for constructing
HST that $O(n^2)$ approximates the given metric. More specifically, the
only part need to be changed is the use of \lemref{small:ball} in
\lemref{l:q:spanner}.  To this end, instead of knowing $\lambda$, we
``guess'' the doubling constant to be $2^i$, increasing $i$ until we ``succeed".
More accurately, in the $i$th iteration, we apply the following sampling
step $2^{3i}$ times: Pick randomly a point $p$ from $P$, and compute
the ball $\Ball(p,r)$ of smallest radius around $p$ containing at
least $n/(2\cdot 2^{3i})$ points.  Next, consider the ball of radius
$\Ball(p,2r)$. If it contains $\leq n/2$ points the algorithm
succeeded, and it stops.  The algorithm is guaranteed to stop when
$i\geq \ceil{\log n}$).  Denote by $\delta=\delta(X)$ the random
value, which is the value of $2^i$ when the algorithm stopped, when
applied to a point set $X\subset \MTR$.

The resulting spanner is a $3n$-approximation \emph{regardless} of the
random bits, and thus the correctness of the \nettree{} algorithm is
guaranteed. We only need to argue about the expected running time for
constructing the HST.  The running time of the HST constructed is
dominated by the spanner construction and the number of edges in it
(see \lemref{l:q:spanner}).  Denote by $\lambda$ the doubling constant
of the metric $\MTR$.

\begin{claim}
    For any $X \subseteq \MTR$, 
    \begin{enumerate}
        \item $\Ex{\delta(X)^{-3}} \geq \lambda^{-3}/16$.
        \item $\Ex{\delta(X)^{3}} = O(\lambda^{3})$.
    \end{enumerate} 
    \clmlab{tedious}
\end{claim}
\begin{proof}
    Consider the algorithm above for computing $\delta(X)$.  Once $i$
    reaches the value $k= \ceil{\log_2 \lambda} $, the probability of
    success on each point sampled is at least $2^{-3 k}$ (by the
    argument in \lemref{small:ball}). Hence the probability of success
    in the $i$th round, $i\geq k$, conditioned on a failure in all
    previous rounds is at least $ 1- (1-2^{-3k})^{2^{3i}} $, which
    means that
    \[ 
    \Ex{\delta(X)^{-3}} \geq 1- (1-2^{-3k})^{2^{3k}} 2^{-3 k} \geq
    (1-1/e)\lambda^{-3}/8. 
    \]

    It also means that $\Pr[\delta \geq 2^{k+i}] \le
    (1-2^{-3k})^{2^{3(k+i-1)}} \leq
    \exp \pth{-\binom{i}{2}}$, and therefore
    \begin{equation*} 
        \Ex{\delta^3}= \sum_{t=1}^\infty \Pr[\delta^3 \geq t] \leq
        2\lambda^3 +\sum_{t=2^{3k}}^\infty
        \exp\left(- \binom{\tfrac{\log t -3k}{3}}{2} \right )
        \leq 2\lambda^3 + O(1)
    \end{equation*}
\end{proof}

We only prove an upper bound on the running time.
Bounding the number of edges is similar. 
Denote by $f(X)$ the running time of the algorithm when applied to
$X\subseteq \MTR$, and let
$g(X)=\Ex{f(X)}$, and $g(n)=\sup_{X\subseteq \MTR,\; |X|=n } g(X)$.

The spanner construction algorithm of \lemref{l:q:spanner} satisfies
\begin{equation}\label{eq:g}
g(X)\leq \Ex{ \max_{\delta(X)^{-3}\leq \alpha \leq 1/2} \pth{ g(\alpha
      |X|) + g((1-\alpha)|X|) +c' \delta(X)^3 n  }}.
\end{equation}
We now prove by induction that $g(n)\leq c \lambda^3 n \ln n$ for some
$c>0$. 
Fix $Y$ to be a subset of $\MTR$ of size $n$
such that $g(Y)=g(n)$. We have
\begin{align*}
    g(n) &\leq
    \Ex{
       \max_{\delta(Y)^{-3}\leq \alpha \leq 1/2} \pth{ \Ex{g(\alpha
             |Y|)} + \Ex{g((1-\alpha)|Y|)} +c'\delta^3n } }\\
    &\leq\Ex{
       \max_{\delta(Y)^{-3}\leq \alpha \leq 1/2} \pth{ c \lambda^3 \alpha
          |Y| \ln ( \alpha |Y| ) + 
          c \lambda^3  (1-\alpha)|Y| \ln \pth{  (1-\alpha)|Y| }  +
c'\delta^3 n}} \\
    &\leq  \Ex{ \MakeBig c \lambda^{3} \delta^{-3} n \ln \pth{\delta^{-3}
          n } +
       c \lambda^3 (1-\delta^{-3}) n \ln \pth{\pth{1-\delta^{-3}}  n}
       + c'\delta^3n } \\
    &\leq  c \lambda^{3} n \cdot 
    \Ex{ \MakeBig  \delta^{-3} \ln \pth{\delta^{-3}
          n}+
       (1-\delta^{-3})  \ln \pth{ \pth{1-\delta^{-3} } n} } 
        +(c''\lambda^3 +d') n \\
    &\leq  c \lambda^{3} n \cdot 
    \Ex{ \MakeBig  \ln \pth{  \pth{1-\delta^{-3}}
          n} } + (c''\lambda^3 +d')n \\
    & \leq c \lambda^3 n \ln n +
    c \lambda^3 n \cdot \Ex{\MakeBig \ln (1-\delta^{-3})} + 
     (c''\lambda^3 +d')n \\
    & \leq  c \lambda^3 n \ln n -
    c \lambda^3 n \cdot \Ex{\MakeBig\delta^{-3}} + (c''\lambda^3 +d')n \\
    & \leq c \lambda^3 n \ln n -
    c \lambda^3 n \cdot \pth{ \lambda^{-3}/16} + (c''\lambda^3 +d')n \\
    &\leq c \lambda^3 n \ln n,
\end{align*}
since $\ln(1-\delta^{-3})\leq - \delta^{-3}$, by
\clmref{tedious}, and for $c>0$ large enough. 
\end{proof}

%-------------------------------------------------------------------------
%-------------------------------------------------------------------------

\section{Concluding Remarks}

In this paper, we show how to efficiently construct 
hierarchical nets for finite spaces with low doubling
dimension, and use it in several applications. 
We believe that this result will have further applications.

Among other things, our fast construction of \WSPD{} implies a near
linear time construction of approximate minimum spanning tree of the
space.  Our fast construction of \nettree{} implies that one can do
$2$-approximate $k$-center clustering in $O(n \log n )$ expected time.

Further transfer of problems and techniques from low dimensional
Euclidean space to low dimensional metrics seems to be interesting. A
plausible example of such a problem is the construction of
$(1+\eps)$-spanners with some additional properties (such as low total
weight or small hop-diameter).  Results of this flavor exist in low
dimensional Euclidean spaces.

It is easy to verify, that for general metric, no HST can be
constructed without inspecting all $\binom{n}{2}$ edges. Indeed,
consider the uniform metric over $n$ points, and change in an
adversarial fashion a single edge to have length $0$.

\subsection{All nearest neighbors.}

The \emph{all nearest neighbor problem} is to compute for a set $P$ of
$n$ points the (exact) nearest neighbor for each point of $p \in P$ in
the set $P \setminus \brc{p}$.  It is known that in low dimensional
Euclidean space this can be done in $O( n \log n)$
time~\cite{c-faann-83,v-aannp-89,ck-dmpsa-95}.  One can ask if a
similar result can be attained for finite metric spaces with low
doubling dimensions. Below we show that this is impossible.

Consider the points $p_1, \ldots, p_n$, where the distance between
$p_i$ and $p_j$, for $i < j$, is either $2^j$ or $2^j+\eps$, for $\eps
< 0.1$. It is easy to verify that this metric has doubling constant at
most three.  We now show that for any deterministic algorithm for
computing all nearest neighbors, there is a metric in the family of the
metrics described above for which the algorithm performs
$\binom{n}{2}$ distance queries.

This claim is proved using an adversarial argument: When the adversary
is queried about the distance between $p_i$ and $p_j$, for $i < j$,
then if not all the distances between $p_1, \ldots, p_{j-1}$ and $p_j$
were specified, the adversary will always return the distance to be $2^j
+ \eps$. The distances $2^j$ would be returned only for the last pair
among the $j-1$ pairs in this set. In particular, for the algorithm to
know what is the closest point to $p_j$, it must perform $j-1$
queries.  Thus, overall, an algorithm doing all nearest neighbors for
$p_1,\ldots, p_n$, will have to perform $\binom{n}{2}$ queries.

A similar asymptotic lower bound can be proved for randomized
algorithms using Yao's principle (here the adversary selects for
each $j$ one index $i_j<j$ at random for which $d(p_{i_j},p_j)=2^j$,
and for the rest of $i\neq i_j$, $i<j$, $d(p_i,j)=2^j+\eps$).  

At this point, it is natural to ask whether one can achieve running
time of $O(n \log( n \Spread(P) ) )$ for the all nearest neighbor
problem. This, however, is straightforward.  Indeed, compute $4$-WSPD
of $P$.  Clearly, if $q$ is a nearest neighbor for $p$, then there is
a pair in the \WSPD{} such that $p$ is the only point on one side, and
the other side contains $q$. Thus, we scan all such unbalanced pairs
(one point on one side, and many points on other side), and compute
the nearest neighbor for each point. Thus, this computes all nearest
neighbors.  As for the running time analysis, consider all such pairs
in distance range $l$ to $2l$, and observe that by a packing argument,
for any node $u$ in the \nettree{}, the number of such \WSPD{} pairs
with $u$ in them is a $2^{O(\dim)}$. In fact, along a path in the
\nettree{}, only a constant number of nodes might participate in such
pairs.  Thus, every point is being scanned $2^{O(\dim)}$ times,
implying that scanning all such pairs takes $2^{O(\dim)}n$ time. There
are $\ceil{\lg(\Spread(P))}$ resolutions, so the overall running time
is $2^{O(\dim)} n \log(n \Spread(P))$.

%-------------------------------------------------------------------------
%-------------------------------------------------------------------------

\subsection*{Acknowledgments}

We thank James Lee, Ken Clarkson, and Alex Slivkins for helpful
correspondence regarding the content of this paper.  James Lee pointed
us to some recent related papers, and suggested the construction of a
doubling measure as an application of the \nettree{}.  The
observations about the all nearest neighbors problem, rose from
discussions with Ken Clarkson. Alex Slivkins pointed out to us an
error in a preliminary version of this paper. Finally, the authors
would like to thank the anonymous referees for their useful comments.

\bibliographystyle{plain} 
\bibliography{lipschitz-bib}

\appendix

\section{Proof of \lemref{correctness}}
\apndlab{net:tree:correct}

Notice that \lemref{g:weak} implies that $(\br_i)_{i\geq 1}$ is
monotone non-increasing sequence, and that $r_i \geq \br_i \geq
r_i/(1+n^{-2}) \geq \frac{4}{5} r_i$.

\medskip

\noindent{}\begin{proofextin}{\emph{\lemref{correctness}}:}
    We prove by induction on $k$ all five assertions together. The
    base case is obvious.  Assume by the induction hypothesis that
    $T^{(k-1)}$ satisfies all the properties above, and we prove it
    for $T^{(k)}$.
    
    \Mparagraph{Property \caseref{find:q}.}  Every point
    inserted during the $l$th {phase} (i.e.,
    a point $p_i$ for which $\ceil{\log_\tbase \br_i}=l$), 
    must have its current
    parent (in $T^{(k)}$) at level $l$. Thus, if $\mlevel{\wu}>l$,
    this means that $c_{p_k}$ was inserted before the current
    {phase}, which means that it is indeed the closest point to
    $p_k$ among $\brc{p_1,\ldots p_h}$.  Otherwise, if
    $\mlevel{\wu}=l$, then
    \[ 
    d_\MTR(\wu,q)\leq d_\MTR(\wu,c_{p_k}) + d_\MTR(c_{p_k},p_k)
    +d_\MTR(p_k,q) \leq 2\cdot \tbase^{l} + (1+n^{-2})\tbase^l +
    (1+n^{-2}) \tau^l \leq
    13 \cdot \tbase^l.
    \] 
    Since $q$ appears before the level $l$ began, either 
    $\mlevel{\parent(q)}>l$ and then $q\in \dRel{\wu}$,
    or $\mlevel{\parent(q)}=l$, but then it must be that
    $\rep_{\parent(q)}=q$, so $\parent(q)\in \dRel{\wu}$.
    Either case $q$ is a representative of a vertex in $\dRel{\wu}$
    which is the same as $\Rel{\wu}$ (in $T^{(k-1)}$).

    \Mparagraph{Property \caseref{bounded:diam}.}  We shall prove it
    both for $p_k$, and the new internal vertex (in case
    \caseref{diff:level1} of the construction).
    Consider first case \caseref{diff:level1} in the construction:
    \[
    d_\MTR(\rep_u,\rep_v)=d_\MTR(\rep_u,q)\leq \tbase^{\mlevel{u}}, 
    \] 
    where the last inequality follows from the induction hypothesis.
    Also,
    \[ 
    d_\MTR(\rep_v,p_k)\leq 2\cdot \tbase^{\mlevel{v}},
    \] 
    and we are done with the first case of the construction.
    
    Case \caseref{same:level1} follows from the definition of $q$,
    and since as argued above, for $u=\parent(q)$,
    $\rep_u=q$.
    
    \Mparagraph{Property \caseref{separation}.}  Fix some $t\in \Re$,
    and let $x$ and $y$ be two vertices for which
    $\max\sbrc{\mlevel{x},\mlevel{y}} < t \leq
    \min\brc{\mlevel{\parent(x)},\mlevel{\parent(y)}}$. If both $x$
    and $y$ are not $p_k$ then the claim follows from the inductive
    hypothesis (even for the new formed internal vertex, since it
    inherits its parent and representative from a previously
    established vertex). Otherwise, assume $x=p_k$.  As $p_k$ is the
    latest addition of leaf to $T$, $d_\MTR(p_k,\rep_y)\geq
    \br_{k-1}\geq \tbase^{l-1}$.  Note that $\mlevel{\parent(p_k)}=l$,
    so $t\leq l$, and we conclude that $d_\MTR(\rep_x,\rep_y)\geq
    \tbase^{t-1}$.

    \Mparagraph{Property \caseref{T:k:net:tree}.}  We next prove that
    $T^{(k)}$ is a \nettree{}. The only non-straightforward claims are
    the packing and covering properties. The covering property follows
    from Property~\caseref{bounded:diam} of this lemma: Let $u=u_1$ be
    a vertex, $v=u_m$ a descendant, and $\langle u_1,\ldots,
    u_m\rangle$ the path between them in $T$, then
    \[
    d_\MTR(\rep_u,\rep_v) \leq \sum_{i=1}^{m-1}
    d_\MTR(\rep_{u_i},\rep_{u_{i+1}}) \leq 2 \sum_{i=1}^{m-1}
    \tbase^{\mlevel{u_i}} \leq 2\sum_{i=1}^{m-1} \tbase^{\mlevel{u_1} -(i-1)}
    \leq \tfrac{2\tau}{\tau-1} \cdot \tbase^{\mlevel{u}} .
    \] 

    The packing property is more delicate. Let $w$ be an arbitrary
    vertex in $T^{(k)}$, and $x \notin P_w$ a point.  We want to prove
    that $d_\MTR(x,\rep_w)\geq \frac{\tbase-5}{2(\tbase-1)}
    \tbase^{\mlevel{\parent(w)}-1}$.  Let $\hx\in T^{(k)}$ be an
    ancestor of $x$ such that $\mlevel{\hx}\leq \mlevel{\parent(w)}-1
    < \mlevel{\parent(\hx)}$.  Applying Property~\caseref{separation}
    with $t=\mlevel{\parent(w)}$, we get that
    $d_\MTR(\rep_{\hx},\rep_{w})\geq \tbase^{\mlevel{\parent(w)}-1}$.

    If $x=\hx$, we are done. Else, if
    $\mlevel{\hx}<\mlevel{\parent(w)}-1$, then, by
    Property~\caseref{bounded:diam}, we have
    \[
    d_\MTR(\rep_w,x)\geq d_\MTR(\rep_w, \rep_{\hx}) - 
    d_\MTR(\rep_{\hx},x) \geq \tbase^{\mlevel{\parent(w)}-1} - 
    \tfrac{2\tbase}{\tbase-1} \cdot \tbase^{\mlevel{\parent(w)}-2} = 
    \tfrac{\tbase-3}{\tbase-1} \cdot \tbase^{\mlevel{\parent(w)}-1}, 
    \]
    and we are done.

    Otherwise, let $\bar{x}\in T^{(k)}$ be an ancestor of $x$ which is
    the child of $\hx$ ($\parent(\bar{x})=\hx$).  If
    $\rep_{\bar{x}}=\rep_{\hx}$ then the preceding argument (where
    $\mlevel{\hx}<\mlevel{\parent(w)}-1$) also applies here, and we
    are done.
    
    Otherwise, we get the following situation:
    $\mlevel{\parent(\bar{x})}=\mlevel{\parent(w)}-1$, and
    $\rep_{\bar{x}}\neq \rep_{\parent(\bar{x})}$, but this can happen
    only if $\bar{x}$ was inserted during level
    $\mlevel{\parent(w)}-1$.  Recall that the algorithm connects
    $\rep_{\bar{x}}$ as a child of vertex in level
    $\mlevel{\parent(w)}-1$ whose representative is the closest point
    among those appearing during the levels greater than
    $\mlevel{\parent(w)}-1$.  Note that both $\rep_{\parent(\bar{x})}$
    and $\rep_w$ inserted in level greater than
    $\mlevel{\parent(w)}-1$, we conclude that
    $d_\MTR(\rep_{\bar{x}},\rep_{\parent(\bar{x})}) \leq
    d_\MTR(\rep_{\bar{x}},\rep_w)$, therefore
    \begin{multline*}
        d_\MTR(\rep_{\bar{x}},\rep_w) 
        \geq \max\sbrc{ d_\MTR(\rep_{\bar{x}},\rep_{\parent(\bar{x})}),
           d_\MTR(\rep_{w},\rep_{\parent(\bar{x})}) -
           d_\MTR(\rep_{\bar{x}},\rep_{\parent(\bar{x})})} \\
        \geq \frac{d_\MTR(\rep_{w},\rep_{\parent(\bar{x})})}{2}
        \geq 0.5 \cdot \tbase^{\mlevel{\parent(w)}-1} .
    \end{multline*}
    Hence, by the covering property, 
    \begin{eqnarray*}
        d_\MTR(x,\rep_w) &\geq& d_\MTR(\rep_{\bar{x}},\rep_w) -
        d_\MTR(\rep_{\bar{x}},x) \geq 0.5 \cdot
        \tbase^{\mlevel{\parent(w)}-1} - \tfrac{2\tbase}{\tbase-1}\cdot
        \tbase^{\mlevel{\parent(w)}-2}\\
        &=&
        \tfrac{\tbase-5}{2(\tbase-1)}\cdot\tbase^{\mlevel{\parent(w)}-1},
    \end{eqnarray*}
    and we are done.

    \Mparagraph{Property \caseref{Rel:dRel}.}  Assume that a new
    vertex $x$ is attached as a child to a vertex $y$. We shall prove
    that our traversing algorithm visits all vertices $w$ for which
    either $w\in \dRel{x}$ or $x\in \dRel{w}$.  Suppose first that $x
    \in \dRel{w}$. Thus, $\mlevel{w}<\mlevel{y}$.  Let $z$ be an
    ancestor of $w$ for which $\mlevel{z}\leq
    \mlevel{y}<\mlevel{\parent(z)}$.  Let $\langle
    z=z_1,\ldots,z_m=w\rangle$ be the path between them in $T$. Then,
    for any $1\leq i\leq m-1$, it holds
    \[
    d_\MTR(\rep_x, \rep_{z_i}) \leq d_\MTR(\rep_x, \rep_{w})+
    d_\MTR(\rep_{z_i},\rep_{w}) \leq 13 \cdot
    \tbase^{\mlevel{z_m}} + \tfrac{2\tbase}{\tbase-1}\cdot 
    \tbase^{\mlevel{z_i}} \leq 13 \cdot
    \tbase^{\mlevel{z_i}} .
    \] 
    Thus $x\in\dRel{z_i}$, for any $2\leq i\leq m$. So if $z=z_1\in
    \Rel{y}$, we are assured that $w=z_m$ will be visited.  Indeed,
    $z\in \dRel{y}$ since,
    \begin{multline*}
        d_\MTR(\rep_y,\rep_z)\leq d_\MTR(\rep_y,\rep_x)+
        d_\MTR(\rep_x,\rep_w) + d_\MTR(\rep_w,\rep_z) \\ \leq 2\cdot
        \tbase^{\mlevel{y}}+ 13 \cdot \tbase^{\mlevel{w}} +
        \tfrac{2\tbase}{\tbase-1}\cdot \tbase^{\mlevel{z}}\leq 2\cdot
        \tbase^{\mlevel{y}}+ 13 \cdot \tbase^{\mlevel{y}-1} +
        \tfrac{2\tbase}{\tbase-1}\cdot \tbase^{\mlevel{y}} \leq 13
        \cdot \tbase^{\mlevel{y}}.
    \end{multline*}
    This means that $z\in \Rel{y}$, by the inductive hypothesis.
    
    Next, we consider the case when $w\in \dRel{x}$.  In this case
    $\mlevel{w} \leq \mlevel{x}<\mlevel{\parent(w)}$ and
    \[ 
    d_\MTR(\rep_w,\rep_y) \leq d_\MTR(\rep_w,\rep_x)+d_\MTR(\rep_x,\rep_y)
    \leq 13\cdot \tbase^{\mlevel{x}}+ 2 \cdot \tbase^{\mlevel{y}} 
    \leq 13\cdot \tbase^{\mlevel{y}}.
    \]
    
    Hence, if $\mlevel{\parent(w)}>\mlevel{y}$ then $w\in \dRel{y}$
    which implies that $w\in \Rel{y}$ by the inductive hypothesis, and
    we are done.

    If $\mlevel{\parent(w)}= \mlevel{y}$ then
    \begin{multline*}
        d_\MTR(\rep_{\parent(w)},\rep_y) \leq 
        d_\MTR(\rep_{\parent(w)},\rep_w)+
        d_\MTR(\rep_w,\rep_x)+d_\MTR(\rep_x,\rep_y) \\
        \leq 2\cdot \tbase^{\mlevel{y}} 
        + 13 \cdot \tbase^{\mlevel{y}-1}+ 2\cdot \tbase^{\mlevel{y}}
        \leq 13  \cdot \tbase^{\mlevel{y}} .
    \end{multline*}
    So, in this case $\parent(w)\in \dRel{y}$, and using the inductive
    hypothesis, we are done.

    We are left with the case $\mlevel{\parent(w)} < \mlevel{y}$.
    In this case
    \begin{multline*}
        d_\MTR(\rep_{\parent(w)},\rep_x) \leq
        d_\MTR(\rep_{\parent(w)},\rep_w)+
        d_\MTR(\rep_w,\rep_x) \\
        \leq 2\cdot \tbase^{\mlevel{\parent(w)}} + 13 \cdot
        \tbase^{\mlevel{\parent(w)}-1} \leq 13 \cdot
        \tbase^{\mlevel{\parent(w)}} .
    \end{multline*}
    So we have that $x\in \Rel{\parent(w)}$.  As was proved above,
    this means that $\parent(w)$ will be visited, and since $x$ is
    added to $\Rel{\parent(w)}$, the algorithm also visits the
    children of $\parent(w)$, and in particular, $w$.
\end{proofextin}

\end{document}